\documentclass[12pt]{article}
\usepackage{amsmath}
\usepackage{graphicx}
\usepackage{amssymb, amsthm}
\usepackage{comment}
\usepackage{algorithm}
\usepackage{algorithmic}

\DeclareFontFamily{U}{mathx}{}
\DeclareFontShape{U}{mathx}{m}{n}{<-> mathx10}{}
\DeclareSymbolFont{mathx}{U}{mathx}{m}{n}
\DeclareMathAccent{\widecheck}{0}{mathx}{"71}

\usepackage[authoryear]{natbib}
\usepackage{url} 
\usepackage{hyperref}
\usepackage{xcolor} 
\usepackage{multirow}
\usepackage{pifont}
\usepackage{soul}

\font\n=cmcsc10
\theoremstyle{remark}
\newtheorem*{lemma*}{Lemma}
\theoremstyle{definition}

\newtheorem{theorem}{\n{Theorem}}[section]

\newtheorem{remark}{\n{Remark}}[section]
\newtheorem{lemma}{\n{Lemma}}[section]

\allowdisplaybreaks
\newcommand{\real}{\mathbb{R}}

\newcommand{\umod}[1]{ \left\| #1 \right\|}

\newcommand{\PP}{\mathbb{P}}

\newcommand{\GG}{\mathcal{G}}
\newcommand{\SC}{\mathcal{S}}
\newcommand{\CC}{\mathcal{C}}

\DeclareMathOperator*{\argmin}{arg\,min}
\newcommand{\blind}{0}

\usepackage{array}
\newcolumntype{H}{>{\setbox0=\hbox\bgroup}c<{\egroup}@{}}
\begin{document}

\def\spacingset#1{\renewcommand{\baselinestretch}%
{#1}\small\normalsize} \spacingset{1}


\if0\blind
{
  \title{\bf A Unified Framework for Community Detection and Model Selection in Blockmodels\\}
  \author{Subhankar Bhadra\hspace{.2cm}\\
    Department of Statistics, The Pennsylvania State University\\
    Minh Tang\\
    Department of Statistics, North Carolina State University\\
    Srijan Sengupta\\
    Department of Statistics, North Carolina State University}
  \maketitle
} \fi

\if1\blind
{
  \bigskip
  \bigskip
  \bigskip
  \begin{center}
    {\LARGE\bf A Unified Framework for Community Detection and Model Selection in Blockmodels\\}
\end{center}
  \medskip
} \fi

\bigskip
\begin{abstract}
Blockmodels are a foundational tool for modeling community structure in networks, with the stochastic blockmodel (SBM), degree-corrected blockmodel (DCBM), and popularity-adjusted blockmodel (PABM) forming a natural hierarchy of increasing generality. While community detection under these models has been extensively studied, much less attention has been paid to the model selection problem, i.e., determining which model best fits a given network. 
Building on recent theoretical insights about the spectral geometry of these models, we propose a unified framework for simultaneous community detection and model selection across the full blockmodel hierarchy. A key innovation is the use of loss functions that serve a dual role: they act as objective functions for community detection and as test statistics for hypothesis testing. We develop a greedy algorithm to minimize these loss functions and establish theoretical guarantees for exact label recovery and model selection consistency under each model.
Extensive simulation studies demonstrate that our method achieves high accuracy in both tasks,  outperforming or matching state-of-the-art alternatives. Applications to five real-world networks further illustrate the interpretability and practical utility of our approach. 
R code for implementing the method is available at \url{https://github.com/subhankarbhadra/model-selection}.

\end{abstract}

\noindent%
\vfill

\newpage
\spacingset{1.5} 
\section{Introduction}
\label{sec:intro}
Blockmodels are a foundational and extensively studied class of statistical models for community structure in networks \citep{newman2010networks, goldenberg2010survey, sengupta2025statistical}.
The stochastic blockmodel (SBM) is the simplest such model, assuming that nodes within a block are stochastically equivalent and that edges form independently based on a block-wise connectivity matrix \citep{lorrain1971structural}. While effective at characterizing community structure in its most basic form, the SBM often falls short when modeling networks that exhibit significant degree heterogeneity or complex interaction patterns.
The degree-corrected blockmodel (DCBM) addresses the degree heterogeneity limitation by introducing node-specific parameters \citep{karrer2011stochastic}. 
A further generalization is the popularity-adjusted blockmodel (PABM) which allows edge probabilities to depend on how popular a node is with respect to each community, thus offering the flexibility to capture complex interaction patterns that arise in real-world networks \citep{senguptapabm}.
The PABM subsumes both the SBM and DCBM as special cases, providing a nested, hierarchical structure for unified modeling and inference across a broad spectrum of network settings \citep{noroozi2021hierarchy, koo2022popularity}.
A rich methodological literature has developed around these models, particularly in the context of community detection. 
This includes spectral clustering and its variants \citep{rohe2011spectral, sussman2012consistent, lei2015consistency, gao2017achieving, sengupta2015spectral, chaudhuri2012spectral}, likelihood-based and pseudo-likelihood methods \citep{zhao2012consistency,amini2013,bickel2009nonparametric}, as well as variational inference and Bayesian techniques \citep{airoldi2009mixed}. More recent works have extended these tools to accommodate the additional complexity of the PABM framework \citep{noroozi2021estimation, noroozi2019sparse, noroozi2021hierarchy,koo2022popularity}.

While community detection has been the primary focus of methodological development under blockmodels,
less attention has been paid to the model selection problem, i.e., 
determining which blockmodel best describes a given network.
This task is essential for informing downstream inference and ensuring that the complexity of the fitted model is appropriate for the observed data.
For example, applying a DCBM or PABM when the simpler SBM suffices can result in unnecessary model complexity, whereas failing to account for degree heterogeneity or complex node popularity patterns may lead to poor fit and misleading interpretations.
Existing model selection methods are designed either to distinguish the SBM from the DCBM or to select the appropriate number of communities under the assumption of an SBM or DCBM.
In early work, \citet{yan2014model} proposed a likelihood-ratio test to distinguish between the SBM and the DCBM.
\citet{lei2016goodness} proposed a goodness-of-fit test for SBMs based on the eigenvalues of the adjacency matrix.
More recently,
\citet{chen2018network}, \citet{li2020network}, and \citet{chakrabarty2025network} have proposed cross-validation techniques to choose between a set of candidate SBMs and DCBMs.

Despite this growing body of work, existing methods suffer from some fundamental limitations. Existing methods are only designed for distinguishing the SBM from the DCBM, and we are not aware of any existing method that incorporates the PABM into the model selection framework. Furthermore, all current methods rely on a two-step procedure: community detection is performed first, and the resulting labels are then used to evaluate model fit through a \textit{separate} test statistic or loss function (typically based on likelihood, spectral gaps, or cross-validation).  
These limitations motivate the need for a unified framework that integrates community detection and model selection while offering theoretical guarantees across a hierarchy of nested blockmodels.

To address these gaps, we propose a unified framework for \textit{simultaneous} community detection and model selection under the full blockmodel hierarchy consisting of the SBM, the DCBM, and the PABM. A central feature of our methodology is the use of model-specific spectral loss functions that serve a \textit{dual} role: they serve both as objective functions for community detection and as test statistics for model selection.
This design leads to a unified workflow that integrates the two inference tasks into a single, coherent pipeline, thus avoiding the current two-step approach.
See Figure~\ref{fig:schema} for a schematic illustration.

Our approach is grounded in two recent and important advances. First, \citet{noroozi2021hierarchy} formalized an elegant, nested hierarchy among the SBM, the DCBM, and the PABM without relying on arbitrary identifiability conditions. In this framework, the SBM corresponds to blockwise constant edge probabilities, the DCBM to blockwise rank-one structure with node-specific degrees, and the PABM to blockwise rank-one matrices derived from node-community popularity vectors. Second, several recent works have studied the spectral structure of the PABM, showing that the latent vectors lie in distinct low-dimensional subspaces \citep{koo2022popularity, noroozi2021estimation}. These important insights enable us to construct spectral loss functions based on subspace projections for each model. To optimize these objective functions, we develop a greedy, computationally efficient algorithm that scales to large networks. Our theoretical results establish consistency guarantees for both community recovery and model selection under each model class. As demonstrated in our numerical experiments, the proposed workflow either outperforms or matches the accuracy of existing state-of-the-art methods in both community detection and model selection tasks.

The rest of the paper is organized as follows. In Section \ref{sec:meth}, we present our unified framework for community detection and model selection.
In Section~\ref{sec:theory}, we establish the theoretical properties of the proposed methodology under each model in the hierarchy. These include exact label recovery guarantees (strong consistency)  for community detection as well as consistency of the model selection procedure, with Type-I error tending to zero and power converging to one for the corresponding hypothesis tests.
In Section~\ref{sec:sim}, we assess the empirical performance of our methodology and compare it against existing state-of-the-art methods.
In Section \ref{sec:data}, we apply the proposed workflow to five real-world networks with community structure and interpret the outcomes. 
Finally, we provide some concluding remarks and discuss potential directions for future research in Section \ref{sec:conc}.
R code for implementing the method is available at \url{https://github.com/subhankarbhadra/model-selection}.

\textit{Notations, models, and setup:} 
Let $A\in\{0,1\}^{n \times n}$ be the adjacency matrix of a simple, undirected network of $n$ nodes with no self-loops, where $A_{ij}\sim \text{Bernoulli}(P_{ij})$ for $1\leq i<j\leq n$ independently. 
We assume that the probability matrix $P$ corresponds to a blockmodel with $K$ communities, where $K$ is known and fixed, i.e., it does not change with $n$. 
Let $\tau_i$ be the community of the $i^{th}$ node, $i\in[n]$, where $[n]$ denotes the set $\{1, \ldots, n\}$.
Let $\GG_k$ be the set of nodes in the $k^{th}$ community, that is, $\GG_k=\{i\colon\tau_i=k\}$ for $k\in[K]$. 

We consider three such blockmodels in this paper.
Under the SBM,
\begin{equation}
    P_{ij}=\Omega_{\tau_i\tau_j},\ 1\leq i<j\leq n,
    \label{sbmdef}
\end{equation}
where $\Omega$ is a ${K\times K}$ symmetric matrix whose entries are in $[0,1]$.
Under the DCBM, 
\begin{equation}
    P_{ij}=\theta_i\,\Omega_{\tau_i\tau_j}\,\theta_j,\ 1\leq i<j\leq n,
    \label{dcbmdef}
\end{equation}
where $\theta_i\in [0,1]$ is a node-specific degree parameter.
Following \cite{lei2015consistency}, we assume the identifiability constraint $\max_{i\in \GG_k}\theta_i=1$ for all $k\in[K]$.
When the $\theta_i$'s are all equal to 1, we get back the SBM as a special case.
Under the PABM, 
\begin{equation}
    P_{ij} = \lambda_{i\tau_j}\, \lambda_{j\tau_i},\ 1\leq i<j\leq n,
    \label{pabmdef}
\end{equation}
where $\Lambda$ is a ${n\times K}$ matrix whose entries are in $[0,1]$.
The $i^{th}$ row of $\Lambda$, $(\lambda_{i1},\ldots,\lambda_{iK})$ can be interpreted as the \textit{popularity} vector of the $i^{th}$ node among the $K$ communities. 
It is easy to see that the DCBM is a special case of the PABM where
$\Lambda_{ik}=\theta_i\sqrt{\Omega_{\tau_ik}},\,i\in[n],\,k\in [K].$

We use standard asymptotic notations, e.g. for sequences $\{a_n\}$ and $\{b_n\}$, $a_n=o(b_n)$ if $a_n/b_n\rightarrow 0$; $a_n=O(b_n)$ if $a_n/b_n$ is bounded above; $a_n=\omega(b_n)$ if $b_n=o(a_n)$; $a_n=\Omega(b_n)$ if $b_n=O(a_n)$; 
$a_n\asymp b_n$ if $a_n=O(b_n)$ and $b_n=O(a_n)$.
We also use the notation $a_n\ll b_n$(resp. $a_n\gg b_n$) which is equivalent to $a_n=o(b_n)$ (resp. $a_n=\omega(b_n)$).
We use $\|\cdot\|$, $\|\cdot\|_F$ and $\|\cdot\|_{2\rightarrow\infty}$ to denote the spectral norm, Frobenius norm and two-to-infinity norm \cite{cape2019twotoinf} of matrices respectively.
We use $\mathrm{diag}(q_1,\ldots,q_r)$ to denote a $r\times r$ diagonal matrix with diagonal elements $q_1,\ldots,q_r$.
For matrices $A$ and $B$, $A\oplus B$ denotes the direct sum of $A$ and $B$.
We say that an event $F_n$ occurs `with high probability' if, for any $c>1$, there exists $C_0>0$ such that $\PP(F_n)\geq 1- C_0\,n^{-c}$. 
\section{Methodology}
\label{sec:meth}

\subsection{Latent positions for SBM and DCBM}
\label{lforsbmdcbm}
For the SBM and the DCBM, we assume that the block probability matrix $\Omega\in[0,1]^{K\times K}$ has rank $K$.
Then, $P$ also has rank $K$.
Let $P=UDU^\top$ be the spectral decomposition of $P$, where $D=\mathrm{diag}(d_1,\ldots,d_K),|d_1|\geq |d_2|\geq\ldots\geq |d_K|>0.$
Let $U_i\in\real^{K}$ be the $i^{th}$ row of $U$.
We define the vectors $\{U_i\}_{1\leq i\leq n}$ as the \textit{latent positions} of the nodes in the network.
It was shown in \citet{lei2015consistency} that, under the SBM, there exist $K$ linearly independent vectors $Y_1,Y_2,\ldots,Y_K\in \real^K$ such that
\begin{equation}
    U_i \,=\, Y_k,\ i\in\GG_k,1\leq k\leq K.
    \label{upos_sbm}
\end{equation}
Similarly under the DCBM, there exist $K$ linearly independent vectors $Y_1,Y_2,\ldots,Y_K\in \real^K$ such that
\begin{equation}
    U_i\,=\,\theta_i\,Y_k,\ i\in\GG_k,1\leq k\leq K.
    \label{upos_dcbm}
\end{equation}
We will define the vectors $\{Y_k\}_{1\leq k\leq K}$ explicitly in Section \ref{sec:theory}. 
From \eqref{upos_sbm} and \eqref{upos_dcbm}, we see that within a community, the $U_i$'s are all equal under the SBM, and the $U_i$'s all lie in a 1-dimensional subspace under the DCBM.
We can estimate $U$ using the Adjacency Spectral Embedding (ASE) method \cite{sussman2012consistent}. Let
$$A=[\widehat{U}\,|\,\widehat{U}^{\perp}]\,[\widehat{D}\oplus \widehat{D}^{\perp}]\,[\widehat{U}\,|\,\widehat{U}^{\perp}]^\top$$
be the spectral decomposition of $A$, where
\begin{align*}
    &\widehat{D}=\mathrm{diag}(\widehat{d}_1,\ldots,\widehat{d}_K),\quad \widehat{D}^{\perp}=\mathrm{diag}(\widehat{d}_{K+1},\ldots,\widehat{d}_n), \quad
    |\widehat{d}_1|\geq|\widehat{d}_2|\geq\ldots\geq|\widehat{d}_K|\geq\ldots\geq |\widehat{d}_n|.
\end{align*}
The ASE of $A$ into $\real^K$ is given by $\widehat{U}$. 

Several papers, including the recent works by \citet{agterberg2025joint} and \citet{xie2021entrywise}, have shown that $\widehat{U}$ is a consistent estimator of $U$ up to an orthogonal transformation under some regularity conditions. 
In particular, it was shown that the maximum row-wise difference, $\max_i\|W\widehat{U}_i - U_i\|$ is small for some $K\times K$ orthogonal matrix $W$.
Although $\widehat{U}$ does not directly estimate $U$, the Euclidean distance between the rows of $\widehat{U}$ is preserved under any orthogonal transformation and our inference procedure only relies on this property.
Therefore, we consider $\widehat{U}_i$ as the $i^{th}$ estimated latent position vector, $1\leq i\leq n$.

\subsection{Latent positions for PABM}
\label{lforpabm}
The PABM can be represented as a special case of a broader class of graph model called the generalized random dot-product graph model (GRDPG), as shown by \citet{koo2022popularity}.
Under the GRDPG model with dimensions $(p,q)$, there exists a $n\times (p+q)$ matrix $X$ such that $P=X\,I_{p,q}\,X^\top$, where $I_{p,q}$ is a diagonal matrix whose first $p$ diagonal entries are all equal to 1, and the remaining $q$ diagonal entries are  $-1$. 
The PABM with $K$ communities  can be represented as a GRDPG model with dimensions $(p=K(K+1)/2,\,q=K(K-1)/2)$, so that $P$ has rank at most $K^2$. 
Let $P=UDU^\top$ be the spectral decomposition of $P$, where $$D=\mathrm{diag}(d_1,\ldots,d_{K^2}), \quad |d_1|\geq|d_2|\geq\ldots\geq|d_{K^2}|>0.$$
Let $U_i\in\real^{K^2}$ be the $i^{th}$ row of $U$. As before, we define the vectors $\{U_i\}_{1\leq i\leq n}$ as the \textit{latent positions} of the nodes in the network. 

Note that for the PABM, the latent positions have dimension $K^2$, unlike latent positions under the SBM and DCBM, which have dimension $K$.
\citet{koo2022popularity} showed that there exist $K$ distinct orthogonal subspaces $\SC_1,\ldots,\SC_K$, each of dimension $K$, such that
\begin{equation*}
    U_i\in \SC_k,\ i\in\GG_k,1\leq k\leq K.
\end{equation*}
We can again estimate $U$ using ASE. 
Let
$A=[\widehat{U}\,|\,\widehat{U}^{\perp}]\,[\widehat{D}\,\oplus\, \widehat{D}^{\perp}]\,[\widehat{U}\,|\,\widehat{U}^{\perp}]^\top$
be the spectral decomposition of $A$, where
$\widehat{D}=\mathrm{diag}(\widehat{d}_1,\ldots,\widehat{d}_{K^2})$, $\widehat{D}^{\perp}=\mathrm{diag}(\widehat{d}_{K^2+1},\ldots,\widehat{d}_n)$, and 
$|\widehat{d}_1|\geq|\widehat{d}_2|\geq\ldots\geq|\widehat{d}_{K^2}|\geq\ldots\geq |\widehat{d}_n|$.
Let $\widehat{U}_i$ be the $i^{th}$ row of $\widehat{U}$.
From \citet{xie2021entrywise}, we have that $U$ is well-approximated by $\widehat{U}$ up to some orthogonal transformation under certain regularity conditions. 
We define $\widehat{U}_i$ as the $i^{th}$ estimated latent position vector, $1\leq i\leq n$.

\subsection{Objective functions and community detection}
\label{sec:objective}
We start with the following observations about the latent positions $\{U_i\}_{1\leq i\leq n}$:
\begin{itemize}
    \item For the SBM, the $U_i$'s are the same within a community, that is, there exist centroids $Y_1,Y_2,\ldots,Y_K\in\real^K$ such that 
    {\setlength{\abovedisplayskip}{0.2em}
     \setlength{\belowdisplayskip}{0.2em}
    \[
    U_i=Y_k,\ i\in\GG_k,1\leq k\leq K.
    \] 
    }
    
    \item For the DCBM, the $U_i$'s lie in a 1-dimensional subspace within a community, that is, there exist rank-1 projection matrices $\Gamma_1,\Gamma_2,\ldots,\Gamma_K\in\real^{K\times K}$ such that 
    {\setlength{\abovedisplayskip}{0.2em}
     \setlength{\belowdisplayskip}{0.2em}
    \[
    U_i=\Gamma_k\, U_i,\ i\in\GG_k,1\leq k\leq K.
    \]}
    
    \item For the PABM, the $U_i$'s lie in a $K$-dimensional subspace within a community, that is, there exist rank-$K$ projection matrices $\Gamma_1,\Gamma_2,\ldots,\Gamma_K\in\real^{K^2\times K^2}$ such that 
    {\setlength{\abovedisplayskip}{0.2em}
     \setlength{\belowdisplayskip}{0.2em}
    \[
    U_i=\Gamma_k\, U_i,\ i\in\GG_k,1\leq k\leq K.\]
    }
\end{itemize}
Based on this observation, we can formulate a unified method for community detection for all three blockmodels. Given the adjacency matrix $A$, we start by obtaining estimates $\{\widehat{U}_i\}$ of the latent positions $\{U_i\}$ using ASE as described in Section \ref{lforsbmdcbm} and \ref{lforpabm}.

Next, for community detection under the SBM, one can simply use the $K$-means algorithm to minimize the objective function
\begin{equation}
  \label{eq:objective_1}
  Q_1(\{t_i\}; \{\widehat{U}_i\}) 
  =  \sum_{i=1}^n  \|\widehat{U}_i - \CC_{t_i}\|^2
  =\sum_{k=1}^K \sum_{i\colon t_i=k} \|\widehat{U}_i - \CC_{k}\|^2 \end{equation}
over all community assignments $\{t_i\}\in [K]^n$, where $\CC_k$ is the $k^{th}$ centroid, defined as the average of $\widehat{U}_i$'s in the $k^{th}$ community, induced by the community assignment $\{t_i\}$. 
This approach is well-grounded in the existing literature, where
applying the $K$-means algorithm to spectral embeddings has been extensively studied \citep{sussman2012consistent,lei2015consistency,sengupta2015spectral}.

For community detection under the DCBM,  
we replace the distance to the $k^{th}$ centroid $\|\widehat{U}_i - \mathcal{C}_{k}\|$ in
Eq.~\eqref{eq:objective_1} with the projection distance onto the subspace spanned by the points in the $k^{th}$ community, i.e., we use the objective function
\begin{equation}
  \label{eq:objective_2}
   Q_2(\{t_i\}; \{\widehat{U}_i\}) =\sum_{k=1}^K \sum_{i\colon t_i=k}\|(I - \Pi_{k}) \widehat{U}_i\|^2,
\end{equation}
where $\Pi_{k}$ is the projection onto the subspace spanned by the best rank-1 approximation to the $\{\widehat{U}_i\}$ belonging to the $k^{th}$ community. 
We note that an alternative approach under the DCBM would be to first
project the points $\widehat{U}_i$ onto the unit sphere before carrying out $K$-means clustering on the projected $\widehat{U}_i$ \citep{lei2015consistency}.

For the PABM, we can use the same objective function as that in Eq.~\eqref{eq:objective_2} but with $\Pi_k$ now being the projection onto the subspace spanned by the best rank-$K$ approximation to the $\{\widehat{U}_i\}$ belonging to the $k$th community, i.e.,
\begin{equation}
  \label{eq:objective_3}
  Q_3(\{t_i\}; \{\widehat{U}_i\}) = \sum_{k=1}^K \sum_{i\colon t_i=k}\|(I - \Pi_{k}) \widehat{U}_i\|^2,
\end{equation}
where $\Pi_k = V_k V_k^{\top}$ and the columns of $V_k$ are the leading (left) singular vectors corresponding to the $K$ largest singular values of some matrix $M_k$ that will be defined subsequently. 

Finally, note that the centroids $\{\CC_k\}$ in Eq. \eqref{eq:objective_1} and the projections $\{\Pi_k\}$ in Eq. \eqref{eq:objective_2} and Eq. \eqref{eq:objective_3} depend on both the community assignments $\{t_i\}$ and latent positions $\{\widehat{U}_i\}$, but we have kept this dependence implicit to avoid notational complexity. In our theoretical analysis, we properly denote the centroids $\{\CC_k\}$ and the projections $\{\Pi_k\}$ as  $\{\CC_k^{(t,\widehat{U})}\}$ and $\{\Pi_k^{(t,\widehat{U})}\}$ respectively, making the dependence explicit. 
We propose the following greedy algorithm for minimizing the objective functions in Eq.~\eqref{eq:objective_2} and Eq.~\eqref{eq:objective_3}.

\begin{enumerate}
    \item Initialization: estimated latent positions $\{\widehat{U}_i\}$, community assignments $\{t_i\}\in[K]^n$ (chosen randomly), tolerance threshold $\tau$, maximum no. of iterations $T$.
    \item This step is slightly different for Eq.~\eqref{eq:objective_2} and Eq.~\eqref{eq:objective_3}.
    \begin{enumerate}
        \item Under the DCBM (Eq.~\eqref{eq:objective_2}), for each community $k$, collect all the points $\widehat{U}_i$ for which $t_i=k$ and call this a matrix $M_k$. Next, find the (left) singular vector $v_k$ corresponding to the largest singular value of $M_k$ and then define $\Pi_k = v_k v_k^{\top}$.
        \item Under the PABM (Eq.~\eqref{eq:objective_3}), for each community $k$, collect all the points $\widehat{U}_i$ for which $t_i=k$ and call this a matrix $M_k$. Next, find the matrix $V_k$ whose columns are the left singular vectors corresponding to the $K$ largest singular values of $M_k$ and then define $\Pi_k = V_k V_k^{\top}$.
    \end{enumerate}
  \item For each point $i$, obtain $t_i^{\text{update}}$ as the community $k$ for which $\|(I - \Pi_k) \widehat{U}_i\|$ is minimized, and compute the step size:
  $$e=\sum_{i=1}^n\|(I - \Pi_{t_i}) \widehat{U}_i\|^2 - \sum_{i=1}^n\|(I - \Pi_{t_i^{\text{update}}}) \widehat{U}_i\|^2.$$
  \item Update community assignments $\{t_i\}=\{t_i^{\text{update}}\}$.
  \item Repeat steps $2$-$4$ until the step size $e$ falls below the threshold $\tau$ or the maximum no. of iterations $T$ is reached.
\end{enumerate}

\begin{remark}
\label{rem:comp_complexity}
    {\bf (Computational complexity)} We now derive the computational complexity of the proposed greedy algorithm. 
Under the PABM, to initialize the algorithm, the estimated latent positions $\widehat{U}_i\in\real^{K^2}$ are obtained via spectral embedding in step 1, which requires $O(\text{nnz}\,(A)K^2 + nK^4)$ time, where $\text{nnz}\,(A)$ denotes the number of non-zero elements in $A$ \citep{halko2011finding}.
Then, in step 2, for each $k$, the matrix $M_k$ can be constructed in $O(n K^2)$ time, and the truncated SVD of $M_k$ can be performed in $O(n K^3)$ time to obtain the left singular vectors corresponding to the $K$ largest singular values, $V_k\in\real^{K^2\times K}$ \citep{halko2011finding}.
Hence, step 2 requires $O(nK^4)$ time.
In step 3, for each $i$, we obtain $t_i^{\text{update}}$ by computing the norms $\|(I - \Pi_k) \widehat{U}_i\|=\|(I - V_kV_k^\top) \widehat{U}_i\|$ for all $k$, which requires $O(K^4)$ time.
The computed norms also provide us the step size $e$.
Therefore, step 3 can be performed in $O(nK^4)$ time.
Step 4 requires only $O(n)$ time.
Finally, steps $2$-$4$ are repeated at most $T$ times.
Therefore, the overall runtime is $O(\text{nnz}\,(A)K^2 + TnK^4)$ time.
Following the same logic, the runtime under the DCBM is  $O(\text{nnz}\,(A)K + TnK^2)$, since in this case $\widehat{U}_i \in \mathbb{R}^K$.
\end{remark}

\subsection{Bootstrap-based model selection}
\label{sec:method_bootstrap}

Building on the objective functions described above, we now propose a two-step testing procedure for model selection. 
Note that if we had access to the true latent positions, i.e., $\{\widehat{U}_i\}=\{U_i\}$, 
all three objective functions $Q_1, Q_2$, and $Q_3$ would be minimized at 0 under the respective (true)  models. 
If we can estimate the latent positions accurately enough, then the minimized objective functions with the estimated latent positions should still be `small' under their respective (true) models. 
Moreover, the minimum of $Q_1$, say $Q_{1}^{\min}$, should be `large' if the underlying model is a DCBM or PABM, as it should not be possible to accurately estimate all the latent positions within a community by a single centroid. 
Similarly, the minimum of $Q_2$, say $Q_{2}^{\min}$, should be `large' when the true model is the PABM, due to the error arising from projecting the PABM latent vectors to a smaller subspace that corresponds to the DCBM.
Thus, we can perform model selection by first observing the value of $Q_{1}^{\min}$ to test whether the underlying model is SBM or DCBM, and if the test is rejected, then we can examine the value of $Q_{2}^{\min}$ to determine whether the model is DCBM or PABM. 
In particular, if $Q_{1}^{\min}$ (resp. $Q_{2}^{\min}$) is small, then we say that there is not strong evidence to reject the SBM in favor of the DCBM (resp. reject the DCBM in favor of the PABM). 

First, we want to test
\begin{equation}
    H_0: A \sim \text{SBM} \; \; \text{ vs. } H_1: A \sim \text{DCBM}.
    \label{test:sbmvsdcbm}
\end{equation}
The test statistic is
$
T_n^{(1)}(A) = Q_1(\{\widehat{\tau}_i\}; \{\widehat{U}_i\}),
$
where
$
\{\widehat{\tau}_i\} = 
\underset{ \{t_i\}\,\in\,[K]^n}{\arg\min}\ Q_1(\{t_i\}; \{\widehat{U}_i\}),
$
and we reject when $T_n^{(1)}(A) > K_\alpha$
for some threshold $K_\alpha$, where $\alpha$ is the level of the test.
The ``correct'' value of $K_\alpha$ is the $(1-\alpha)$-quantile of the sampling distribution of $T_n^{(1)}(A)$ under the null.
However, this sampling distribution is challenging to formulate. 
Therefore, we propose a parametric bootstrap strategy to estimate the threshold $K_\alpha$.
Given a network $A$, we fit an SBM by using $Q_1$ and obtain an estimate of $P$, say $\widehat{P}$.
Next, we generate $R$ replicates $A^*_1, \ldots, A^*_R \sim \text{Bernoulli}(\widehat{P})$, and compute $T_n^{(1)}(A^*_1), \ldots, T_n^{(1)}(A^*_R)$.
The $p$-value is given by 
$
\frac{1}{R} \sum_{r=1}^R \mathbb{I}(T_n^{(1)}(A^*_r) \geq T_n^{(1)}(A)),
$
where $\mathbb{I}(.)$ is the indicator function.

If the test in \eqref{test:sbmvsdcbm} is rejected, then we test
\begin{equation}
    H_0: A \sim \text{DCBM} \; \; \text{ vs. } H_1: A \sim \text{PABM}.
    \label{test:dcbmvspabm}
\end{equation}
The test statistic is
$
T_n^{(2)}(A) = Q_2(\{\widehat{\tau}_i\}; \{\widehat{U}_i\}),
$
where
$
\{\widehat{\tau}_i\} = 
\underset{ \{t_i\}\,\in\,[K]^n}{\arg\min}\ Q_2(\{t_i\}; \{\widehat{U}_i\}),
$
and we reject when $T_n^{(2)}(A) > K_\alpha$
for some threshold $K_\alpha$, where $\alpha$ is the level of the test. Again, $K_{\alpha}$ can be estimated using the bootstrap procedure as before.
The rejection thresholds for both tests are given in the theorems in the next section.

\subsection{Summary of Proposed Methodology and Workflow}

We conclude this section with a concise summary of the proposed methodology for simultaneous community detection and model selection. A key innovation of our approach is its \textit{unified framework} that integrates both tasks using a common spectral embedding pipeline and model-specific objective functions. This is in contrast to standard two-step approaches in the literature, which first estimate communities and then use them to compute a separate goodness-of-fit or likelihood-based loss function for model selection \citep{li2020network, lei2016goodness, chakrabarty2025network}. Our framework avoids this decoupling by using the {same objective functions} for clustering and hypothesis testing.
The full workflow is as follows:

\begin{enumerate}
    \item \textbf{Adjacency Spectral Embedding:} Given a network adjacency matrix \( A \in \{0,1\}^{n \times n} \), carry out its spectral decomposition to find the top $d$ eigenvalues and eigenvectors,
    \[
    A = [\widehat{U} \mid \widehat{U}^\perp]
    \begin{bmatrix}
    \widehat{D} & 0 \\
    0 & \widehat{D}^\perp
    \end{bmatrix}
    [\widehat{U} \mid \widehat{U}^\perp]^\top,
    \]
    and define the latent positions as \( \widehat{U}_i \in \mathbb{R}^d \), the \(i\)th row of \( \widehat{U} \), where \( d = K \) for SBM/DCBM and \( d = K^2 \) for PABM.

    \item \textbf{SBM vs. DCBM:}  Implement K-means to minimize the objective functions \( Q_1 \), and let \( Q_1^{\min} \) denote the minimized value.
     Test
       $ H_0: A \sim \mathrm{SBM} \quad \text{vs.} \quad H_1: A \sim \mathrm{DCBM}$
    using the test statistic \( T_n^{(1)} = Q_1^{\min} \).
    
    If the test is rejected, proceed to the next step. If the test is not rejected, exit the workflow and conclude that the correct model is SBM and the estimated community structure is given by the minimizer of \( Q_1 \).
    Theorems \ref{thm:q1_sbm} and \ref{thm:q1_dcbm} provide theoretical guarantees of exact label recovery and model selection consistency in this case.

     \item \textbf{DCBM vs. PABM:} Implement the greedy algorithm from Section \ref{sec:objective} to minimize the objective function \( Q_2 \), and let  
    \( Q_2^{\min} \) denote the minimized value.
    Test
       $ H_0: A \sim \mathrm{DCBM} \quad \text{vs.} \quad H_1: A \sim \mathrm{PABM}$
    using the test statistic \( T_n^{(2)} = Q_2^{\min} \).
    
    If the test is rejected, proceed to the next step. If the test is not rejected, exit the workflow and conclude that the correct model is DCBM and the estimated community structure is given by the minimizer of \( Q_2 \).
    Theorems \ref{thm:q2_dcbm} and \ref{thm:q2_pabm_old} provide theoretical guarantees of exact label recovery and model selection consistency in this case.

    \item \textbf{PABM:} Implement the greedy algorithm from Section \ref{sec:objective} to minimize the objective function \( Q_3 \). 
        Conclude that the correct model is PABM and the estimated community structure is given by the minimizer of \( Q_3 \).
    Theorem \ref{thm:q3_pabm} provides a theoretical guarantee of exact label recovery in this case.
\end{enumerate}

\begin{figure}[ht]
    \centering
    \includegraphics[trim={10cm 7cm 0 4cm}, clip,height=0.38\textheight]{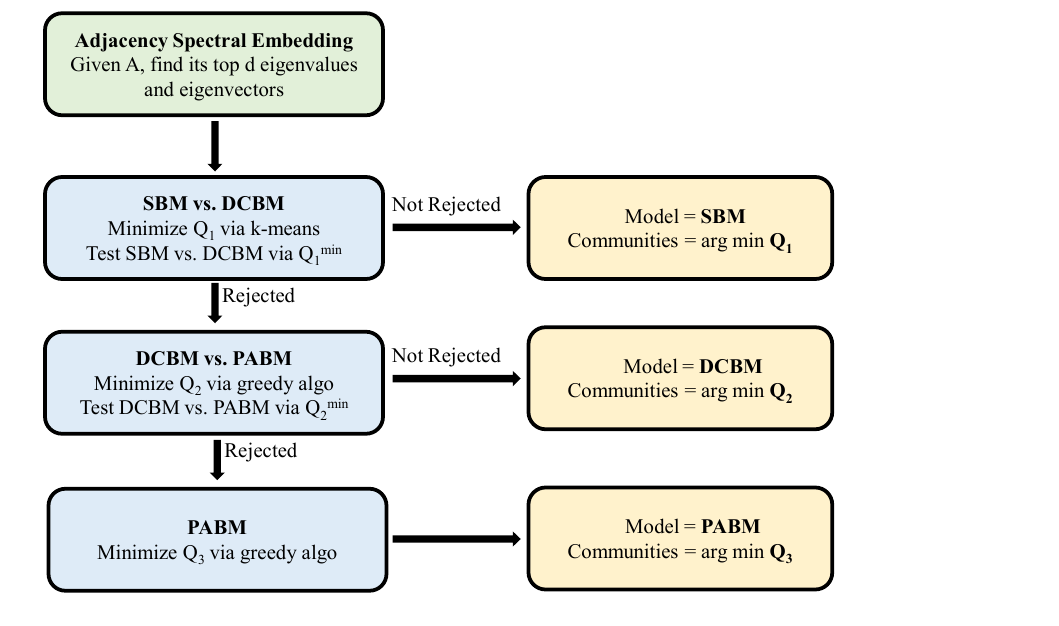}
    \caption{\small Visual schematic of the proposed workflow}
    \label{fig:schema}
\end{figure}

As mentioned within each step, the theoretical results in Section \ref{sec:theory} provide statistical guarantees for model selection consistency and exact label recovery under each scenario.
Figure \ref{fig:schema} provides a visual schematic of the proposed workflow. 
The steps are described in the form of an algorithm in Algorithm \ref{algo1}.

\begin{remark}
    {\bf(Selecting $K$)} In this paper, we assume that $K$ is known. 
    However, in practice, $K$ is often unknown. In such cases, we recommend estimating $K$ under the \textit{current} {null model} before minimizing the corresponding loss function.
    There has been a growing body of work on estimating $K$ in blockmodels, which can be used for this purpose \citep{lei2016goodness,chen2018network,li2020network,le2022estimating, chakrabarty2025network}.
   We note that the output from the unified framework will be sensitive to the statistical uncertainty inherent in estimating $K$.
    However, since existing estimators are known to be
statistically consistent, the unified method will remain consistent
after accounting for estimation error. A thorough investigation of the various estimators for $K$ and their impact on subsequent inference, for both simulated and real-world data, is beyond the
scope of the present paper, but we view it as an important direction for future work.
\end{remark}

\begin{algorithm}[ht]
\caption{Unified Framework for Community Detection and Model Selection}
\label{algo1}
\begin{algorithmic}[1]
\REQUIRE Adjacency matrix \( A \in \{0,1\}^{n \times n} \), no. of communities \( K \), level of significance \(\alpha\)
\ENSURE Estimated community labels \( \{\widehat{\tau}_i\} \), selected model

\STATE Perform adjacency spectral embedding to compute top \( K \) eigenvectors

\STATE Compute \( Q_1^{\min} = \min_{\{t_i\}} Q_1(\{t_i\}; \{\widehat{U}_i\}) \) using K-means

\STATE Perform parametric bootstrap to test 
$H_0: A \sim \mathrm{SBM} \quad \text{vs.} \quad H_1: A \sim \mathrm{DCBM}$

\IF{p-value \( \geq \alpha \)}
    \RETURN \( \{\widehat{\tau}_i\} = \arg\min Q_1(\cdot; \{\widehat{U}_i\}) \),   {Model: SBM}
\ENDIF

\STATE Compute \( Q_2^{\min} = \min_{\{t_i\}} Q_2(\{t_i\}; \{\widehat{U}_i\}) \) using greedy algorithm

\STATE Perform parametric bootstrap to test
$H_0: A \sim \mathrm{DCBM} \quad \text{vs.} \quad H_1: A \sim \mathrm{PABM}$

\IF{p-value \( \geq \alpha \)}
    \RETURN \( \{\widehat{\tau}_i\} = \arg\min Q_2(\cdot; \{\widehat{U}_i\}) \),   {Model: DCBM}
\ENDIF

\STATE Perform adjacency spectral embedding to compute top \( K^2 \) eigenvectors

\STATE Compute \( Q_3^{\min} = \min_{\{t_i\}} Q_3(\{t_i\}; \{\widehat{U}_i\}) \) using greedy algorithm

\RETURN \( \{\widehat{\tau}_i\} = \arg\min Q_3(\cdot; \{\widehat{U}_i\}) \),  {Model: PABM}

\end{algorithmic}
\end{algorithm}

\section{Theoretical results}
\label{sec:theory}

In this section, we establish the theoretical foundations of our unified framework for community detection and model selection under the SBM, DCBM, and PABM. Under each model, we prove exact label recovery for community detection as well as consistency of Type-1 error and power. 
We first state a result for the Frobenius and $2 \to \infty$ norm estimation of the latent positions $U$. 
\begin{lemma}
\label{lem:estimation_U}
Let $A$ be an edge-independent random graph with edge
probability matrix $P\in[0,1]^{n\times n}$, where $P$ has rank $r$ and $r$ is a constant independent of $n$.  Let $\widehat{U}$
and $U$ be $n \times r$ matrices whose columns are the leading eigenvectors of $A$ and $P$ respectively. 
Let $\delta(P) := \max\limits_{i\in[n]} \sum_{j=1}^n P_{ij}$ and $d_{r}(P)$ denote the maximum expected degree and smallest (in magnitude) non-zero eigenvalue of $P$, respectively. 
Suppose that $\delta(P)=\Omega(\log n)$ for all $n\in\mathbb{N}$ and $|d_{r}(P)| \geq c_0\,\delta(P)$ for some constant $c_0$. 
Denote by $\mathbb{O}_r$ the set of $r \times r$ orthogonal matrices. Then for any constant $c > 0$, there exist constants $C > 0$ and $n_0 > 0$, both possibly depending on $c$, such that for all $n \geq n_0$ we have \begin{gather}
  \label{eq:bound_Uhat_Frob}
  \min_{W \in \mathbb{O}_r} \|\widehat{U}W  - U\|_{F}^2 \leq \frac{C}{\delta(P)}, \\[.1cm]
  \label{eq:bound_Uhat_2_to_inf}
\min_{W \in \mathbb{O}_r} \|\widehat{U}W - U\|_{2 \to \infty} \leq
  \frac{C \log^{\frac{1}{2}}{n}}{n^{\frac{1}{2}} \delta(P)^{\frac{1}{2}}},
\end{gather}
with probability at least $1 - n^{-c}$. 
\end{lemma}

Eq.~\eqref{eq:bound_Uhat_Frob} follows from the Davis-Kahan theorem and standard matrix concentration bounds for $\|A - P\|$ (see e.g., \cite{oliveira2009concentration,bandeira2016sharp,tropp2012user}), while
Eq.~\eqref{eq:bound_Uhat_2_to_inf} is an adaptation/simplification of Theorem~3.2 in \cite{xie2021entrywise}
to the setting of the current paper (a similar result is also provided in
\cite{cape2019signal} but with a slightly worse lower bound condition for $\delta(P)$). 
As $U$ (resp. $\widehat{U}$) is not unique unless the $r$ largest eigenvalues of $P$ (resp. $A$) are distinct, these
Frobenius and $2 \to \infty$ norm bounds involve minimization over orthogonal matrices $W$ to align the subspaces for $\widehat{U}$ and $U$. For ease of exposition, we omit the dependency on this alignment in the subsequent discussion as it has no impact on the theoretical results; more specifically,
our inference procedures only depend on the Euclidean distance between the rows of $\widehat{U}$ and thus yield the same performance when applied to $\widehat{U} W$ for any arbitrary orthogonal matrix $W$. 

\subsection{Stochastic blockmodel}

Let us consider an SBM with parameter $\Omega$ as defined in \eqref{sbmdef}. 
By Lemma 3.1 of \cite{lei2015consistency}, there exists a $K \times K$ matrix $H$ with orthonormal rows such that
\begin{equation}
    \label{eq:def_U_H}
    U_{i}= n_{\tau_i}^{-\frac{1}{2}} H_{\tau_i}
\end{equation}
for all $i \in [n]$, where $n_{k} = |\mathcal{G}_k|$ for all $k \in [K]$, and $U_i$ and
$H_{k}$ are the $i^{th}$ and $k^{th}$ row of $U$ and $H$, respectively.
Define $n_{\min}=\min_{k} n_k,\,n_{\max}=\max_{k} n_k$.
Let $\delta(P) = \max_{i} \sum_{j=1}^n P_{ij}$ denote the maximum expected degree.
The assumptions we are going to consider are
\begin{itemize}
  \item[\textbf{A1.}]\ The communities are balanced, that is, there exists a constant $c>0$ such that
    $$\frac{n}{c K}\leq n_k\leq \frac{c\,n}{K},\ \text{for all }k\in[K].$$
  \item[\textbf{A2.}]\ $d_K\geq c_0\,\delta(P)$ for some constant $c_0>0$, $\delta(P) = \omega(\log n)$. 
\end{itemize}
\textbf{A1} is the balanced communities assumption, which says that the community sizes are of the same order.
\textbf{A2} is a lower bound on the sparsity of the network.

We are now ready to state the theoretical results under the SBM. The following theorem shows that when the network is generated from an SBM, minimizing the objective function \( Q_1 \) using $K$-means on spectral embeddings leads to exact recovery of the true community labels with high probability. Moreover, the minimized objective function, \( Q_1^{\min} \), remains small under the SBM with probability going to 1. 
We note that strong consistency under the SBM is not a new result in itself \citep{sussman2012consistent,lei2015consistency}, but the additional result proving an upper bound on the objective function is new and critical for model selection, since it validates the use of \( Q_1^{\min} \) as a test statistic for distinguishing the SBM from more complex models.

\begin{theorem}
\label{thm:q1_sbm}
    Suppose $A$ is the adjacency matrix of a network from the SBM with parameter $\Omega$ as defined in Eq. \eqref{sbmdef}. Let Assumptions \textbf{A1}-\textbf{A2} hold.
    Let $\{\widehat{\tau}_i\}$ be a $(1+\epsilon)$-approximate solution to $Q_1(\{t_i\};\{\widehat{U}_i\})$, that is, for a given $\epsilon>0$,
    \begin{equation}
        Q_1(\{\widehat{\tau}_i\}; \{\widehat{U}_i\}) \leq (1+\epsilon)\,\underset{ \{t_i\}\,\in\,[K]^n}{\min}\ Q_1(\{t_i\}; \{\widehat{U}_i\}).
        \label{main.apx.sol.1}
    \end{equation}
    Then,
    $$Q_1(\{\widehat{\tau}_i\};\{\widehat{U}_i\}) = O(\delta(P)^{-1})$$
    with high probability. Furthermore, there exists a bijection $\sigma:[K]\mapsto[K]$ such that 
    $$\sum_{i=1}^n \mathbb{I}(\widehat{\tau}_i \not = \sigma(\tau_i)) = 0$$
    with high probability, i.e., $\widehat{\tau}$ achieves exact recovery of $\tau$.
\end{theorem}

\subsection{Degree-corrected blockmodel}
We now consider a DCBM with parameters $\Omega$ and $\{\theta_i\}_{1\leq i\leq n}$, as defined in \eqref{dcbmdef}.
Define $\theta_{\min}=\min_{i}\theta_i,\,\theta_{\max}=\max_{i}\theta_i$.
Let $\phi_k \in \mathbb{R}^n$ be a vector which agrees with $\theta$ on $\GG_k$ and 0 elsewhere. Let $\widetilde{\phi}_k=\phi_k/\umod{\phi_k}$ and $\widetilde{\theta}=\sum_{k=1}^K \widetilde{\phi}_k$.
By Lemma 4.1 of \cite{lei2015consistency}, there exists a $K \times K$ matrix $H$ with orthonormal rows such that 
\begin{equation}
\label{eq:def_U_H_dcbm}
    U_{i}\,=\,\widetilde{\theta}_i\, H_{\tau_i}
\end{equation}
for all $i \in [n]$, where $H_k$ is the $k^{th}$ row of $H$. 

The next theorem is regarding the behavior of $Q_1$ under the DCBM.
It shows that when the network is generated from a DCBM with sufficient degree heterogeneity, the minimized objective function, $Q^{\min}_1$, is greater than $1/\delta(P)$  with high probability for any choice of community assignments.
This result, in conjunction with the second part of Theorem 3.1, ensures the asymptotic power of the SBM vs. DCBM test and validates the rejection of the SBM in favor of the DCBM as long as  there is sufficient variation among the $\theta_i$'s. 

\begin{theorem}
\label{thm:q1_dcbm}
Let $A$ be the adjacency matrix of a network from the DCBM with parameters $\Omega$ and $\{\theta_i\}_{1\leq i\leq n}$ as defined in Eq.~\eqref{dcbmdef}. 
Let Assumptions \textbf{A1}-\textbf{A2} hold. Furthermore, suppose $\theta_{\max}/\theta_{\min} = O(1)$ and
\begin{equation}
    \frac{1}{n_{\max}} \sum_{k=1}^K \sum_{i\colon\tau_i=k}(\theta_i - \bar{\theta}_{k})^2\gg \frac{1}{\delta(P)}, \quad 
    \text{ where }\bar{\theta}_{k}=\frac{1}{n_k}\sum_{i\colon \tau_i=k}\theta_i. 
    \label{thetavarcond}
\end{equation}
Then for any community assignment $\{t_i\}$,  with high probability we have
$$Q_1(\{t_i\};\{\widehat{U}_i\}) \gg 1/\delta(P).$$
\end{theorem}

\begin{remark}
    Theorem \ref{thm:q1_dcbm} states that if there is sufficient heterogeneity among the degree parameters $\{\theta_i\}$ generating the DCBM network in the form of Eq. \eqref{thetavarcond}, then the test based on $Q_1$ has an asymptotic power of 1. 
    If the entries of $\Omega$ scales with a sparsity parameter $\rho_n$, then $\delta(P)\asymp n\rho_n$, and Eq. \eqref{thetavarcond} reduces to
    \begin{equation}
        \frac{1}{n_{\max}} \sum_{k=1}^K \sum_{i\colon\tau_i=k}(\theta_i - \bar{\theta}_{k})^2\gg 1/(n\rho_n).
        \label{thetavarcond2}
    \end{equation}
    Theorem 3.5 of \cite{lei2016goodness} imposes a similar restriction on the heterogeneity of the degree parameters for the goodness-of-fit test.
    Under the assumptions of Theorem \ref{thm:q1_dcbm}, Lei's test has asymptotically power 1 provided there exists a community $k\in[K]$ such that 
    \begin{equation}
        \frac{1}{n_{\max}}\min_{u\,\in\,\real^{n_{\kern -0.1em k}}} \left\{\sum_{i:\,\tau_i=k}(\theta_i-u)^2: u\ \text{has at most $K$ distinct values}\right\} \gg 1/(n^{\frac{1}{2}}\rho_n).
        \label{thetavarcond_lei}
    \end{equation}
    Clearly, Eq. \eqref{thetavarcond2} is a much weaker condition than Eq. \eqref{thetavarcond_lei}, which implies that the proposed test is asymptotically more powerful than the goodness-of-fit test by \citet{lei2016goodness}.
\end{remark}

Our next result shows that minimizing $Q_2$ leads to exact label recovery of $\tau$ under a DCBM.
We note that there are existing community detection methods in the literature that also achieve exact label recovery.
This theorem shows that the proposed community detection achieves comparable accuracy.
Furthermore, it also shows that 
$Q^{\min}_2$
is small under the DCBM, justifying its role as a test statistic for model selection.

\begin{theorem}
\label{thm:q2_dcbm}
    Suppose $A$ is the adjacency matrix of a network from the DCBM with parameters $\Omega$ and $\{\theta_i\}_{1\leq i\leq n}$ as defined in Eq. \eqref{dcbmdef}. Let Assumptions \textbf{A1}-\textbf{A2} hold and that $\theta_{\max}/\theta_{\min} = O(1)$. Let $\{\widehat{\tau}_i\}$ be a $(1+\epsilon)$-approximate solution to $Q_2(\{t_i\};\{\widehat{U}_i\})$, that is, for a given $\epsilon>0$,
    \begin{equation}
        Q_2(\{\widehat{\tau}_i\}; \{\widehat{U}_i\}) \leq (1+\epsilon)\,\underset{ \{t_i\}\,\in\,[K]^n}{\min}\ Q_2(\{t_i\}; \{\widehat{U}_i\}).
        \label{main.apx.sol.2}
    \end{equation}
    Then
    $Q_2(\{\widehat{\tau}_i\};\{\widehat{U}_i\}) = O(\delta(P)^{-1})$ {with high probability}, 
    and there exists a bijection $\sigma:[K]\mapsto[K]$ such that 
    $\sum_{i=1}^n \mathbb{I}(\widehat{\tau}_i \not = \sigma(\tau_i)) = 0$
    with high probability, i.e., $\widehat{\tau}$ achieves exact recovery of $\tau$. 
\end{theorem}

\subsection{Popularity-adjusted blockmodel}
Let us consider a PABM with popularity parameters $\Lambda$ as defined in Eq.~\eqref{pabmdef}.
Let $UDU^\top$ be the eigendecomposition of $P$, where
$$D=\mathrm{diag}(d_1,\ldots,d_{K^2}),|d_1|\geq| d_2|\geq\ldots\geq|d_{K^2}|>0.$$
We now note a few simple properties of $U$ that is essential to the subsequent discussion. 
First assume, without loss of generality, that the rows of $U$ are ordered in increasing order of the true community assignments $\tau$, i.e., $\tau_i \leq \tau_j$ for all $i \leq j$. Next, for any $(k,\ell) \in [K] \times [K]$, let
$\lambda^{(k \ell)} = (\lambda_{i \ell} \colon \tau_i = k)$ denote the vector in $\mathbb{R}^{n_k}$
whose elements are the affinities toward the
$\ell^{th}$ community of all vertices in the $k^{th}$ community. Define
$$\Lambda^{(k)} = [\lambda^{(k1)} \mid \lambda^{(k2)} \mid \dots \mid \lambda^{(kK)}] \in \mathbb{R}^{n_k \times K}, \quad X = \begin{bmatrix} \Lambda^{(1)} & 0 & 0 & \dots & 0 \\ 0 & \Lambda^{(2)} & 0 & \dots & 0 \\
  \vdots & \vdots & \ddots & \vdots & \vdots \\ 0 & 0 & \dots & 0 & \Lambda^{(K)} \end{bmatrix} \in \mathbb{R}^{n \times K^2}.$$
Then from the proof of Theorem~2 in \citet{koo2022popularity}, we have $UU^{\top} = X(X^{\top} X)^{-1} X^{\top}$, and hence, $U = X(X^{\top} X)^{-\frac{1}{2}} W$ for some orthogonal matrix $W$. Let $Z = X(X^{\top} X)^{-\frac{1}{2}}$. Then $Z$ have the same block diagonal structure as $X$, i.e, 
\[ Z = \begin{bmatrix} Z^{(1)} & 0 & 0 & \dots & 0 \\ 0 & Z^{(2)} & 0 & \dots & 0 \\ \vdots & \vdots & \ddots & \vdots & \vdots \\ 0 & 0 & \dots & 0 & Z^{(K)} \end{bmatrix}, Z^{\top}Z = \begin{bmatrix} Z^{(1)\top} Z^{(1)} & 0 & 0 & \dots & 0 \\
  0 & Z^{(2)\top} Z^{(2)} & 0 & \dots & 0 \\
  \vdots & \vdots & \ddots & \vdots & \vdots \\
  0 & 0 & 0 & \dots & Z^{(K)\top} Z^{(K)} \end{bmatrix} = I_{K^2}, \]
where $Z^{(k)}$ is a $n_k \times K$ matrix for all $k \in [K]$. 
If a matrix $\mathcal{P}$ is symmetric idempotent, then
$W \mathcal{P} W^{\top}$ is symmetric idempotent for any orthogonal $W$
and $\|(I - \mathcal{P}) U_i\| = \|(I - W \mathcal{P} W^{\top}) (W U_i)\|$ for all $i$. Therefore,
we can assume without loss of generality that $U = Z$ so that
$\sum_{i \colon \tau_i = k} \|U_i\|^2 = \|Z^{(k)}\|_F^2 = \mathrm{tr} [Z^{(k)\top} Z^{(k)}] = K$. 

We make the following assumptions for the node popularity vectors $\lambda_i = (\lambda_{i1}, \dots, \lambda_{iK})$:
\begin{itemize}
    \item[\textbf{B1.}] The communities are balanced, that is, there exists $c>0$ such that
    $$\frac{n}{c\,K}\leq n_k\leq \frac{c\,n}{K},\ \text{for all }1\leq k\leq K.$$
    \item[\textbf{B2.}]\ $|d_{K^2}|\geq c_0\,\delta(P)$ for some constant $c_0>0$, $\delta(P) = \omega(\log n)$.
    \item[\textbf{B3.}] There exists a sequence $\{\rho_n\}$ such that $\|\Lambda\|_{2\rightarrow\infty}=\rho_n^{\frac{1}{2}}$, and there exists a constant $\gamma_1 > 0$ such that
    $\|\lambda_{i}\| \geq \gamma_1\,\rho_n^{\frac{1}{2}}$ for all $i \in [n]$.
    \item[\textbf{B4.}] There exists a constant $\gamma_2 > 0$ such that for any $k \in [K]$, and any
    subset of nodes $\mathcal{S}$ from community $k$ with $|\mathcal{S}|\geq n_k/K$, we have
    \begin{equation*}
        \sigma_{\min}\left(\sum_{i\,\in\,\mathcal{S}} \lambda_i \lambda_i^{\top}\right)\, \geq\, \gamma_2\,\frac{n_k\rho_n}{K},
        \label{pabmcond_1}
    \end{equation*}
     where $\sigma_{\min}(\cdot)$ denote the smallest singular value of a matrix.

\end{itemize}

\textbf{B1} is the balanced communities assumption, which says that the community sizes are of the same order.
\textbf{B2} is a lower bound on the sparsity of the network.
\textbf{B3} is a standard homogeneity condition on the scale of the latent position vectors.
The quantity $\rho_n$ introduced in \textbf{B3} can be interpreted as a sparsity parameter, and \textbf{B2} implies that $n\rho_n$ must be $\omega(\log n)$.
\textbf{B4} states that any sufficiently large collection of vertices from a given community $k$
has latent positions that cover (in volume) a non-negligible region of $\Lambda^{(k)}$.


Our next result derives the behavior of the objective function \( Q_2 \) under the PABM. Recall that, when the true model is DCBM, the latent positions lie in distinct 1-dimensional subspaces, one for each community. However, under the PABM, the true latent positions reside in a higher-dimensional space, specifically in the span of the top \( K^2 \) eigenvectors of the population matrix \( P \). To study the behavior of \( Q_2 \) under the PABM, we must therefore analyze the properties of the \( K \)-dimensional vector embeddings derived from the top \( K \) eigenvectors of \( P \), which no longer coincide with the true latent positions. This is different from the SBM vs.\ DCBM case, where the latent positions are nested. Assuming \( |d_K| > |d_{K+1}| \), let \( \widetilde{U} \) denote the matrix of the first \( K \) eigenvectors of \( P \), and \( \widehat{\widetilde{U}} \) be its empirical counterpart. The following theorem shows that, under the PABM, if the rows of \( \widetilde{U} \) span higher-dimensional subspaces across communities, then \( Q_2 \) exceeds \( 1/\delta(P) \) for any community assignment. This guarantees that the DCBM vs.\ PABM test achieves power tending to one under the PABM alternative.

\begin{theorem} 
\label{thm:q2_pabm_old}
    Suppose $A$ is the adjacency matrix of a network from the PABM with parameter $\Lambda$ as defined in \ref{pabmdef}. 
    Let Assumptions \textbf{B1}-\textbf{B2} hold.
    Suppose that there exists at least one community $k\in[K]$ such that for any subset of nodes $\mathcal{S}$ of size at least $n_k/K$ from that community,
    \begin{equation}
        \sigma_2(\widetilde{U}_{\mathcal{S}}) \gg 1/\delta(P),
        \label{pabmcond_2}
    \end{equation}
    where $\widetilde{U}_{\mathcal{S}}\in\real^{K\times |\mathcal{S}|}$ is the matrix consisting of columns $\widetilde{U}_i,\,i\in\mathcal{S}$, and $\sigma_2(.)$ denotes the second largest singular value of a matrix.
        Then for any community assignment $\{t_i\}$, we have
    $Q_2(\{t_i\}; \{\widehat{\widetilde{U}_i}\}) \,\gg\, 1/\delta(P)$
    with high probablity. 
\end{theorem}
The condition \eqref{pabmcond_2} states that for any sufficiently large collection of vertices from a given community $k$,
the $K$-dimensional embeddings $\{\widetilde{U}_i\}$ cannot be well approximated by a 1-dimensional subspace in $\real^K$.

Our final theorem shows that the proposed objective function $Q_3$, which clusters spectral embeddings using rank-$K$ projections, yields exact community recovery under the PABM.
It validates the final step of the model selection workflow in Figure \ref{fig:schema}, ensuring that both clustering and model selection are consistent under the PABM.

\begin{theorem}
\label{thm:q3_pabm}
  Suppose $A$ is the adjacency matrix of a network from the PABM with parameters $\Lambda$ as defined in
  Eq.~\eqref{pabmdef}. 
    Let Assumptions \textbf{B1}-\textbf{B4} hold and $\{\widehat{\tau}_i\}$ be a $(1+\epsilon)$-approximate solution to $Q_3(\{t_i\};\{\widehat{U}_i\})$, that is, for a given $\epsilon>0$,
    \begin{equation}
        Q_3(\{\widehat{\tau}_i\}; \{\widehat{U}_i\}) \leq (1+\epsilon)\,\underset{ \{t_i\}\,\in\,[K]^n}{\min}\ Q_3(\{t_i\}; \{\widehat{U}_i\}).
        \label{main.apx.sol.3}
    \end{equation}
    Then   
        $Q_3(\{\widehat{\tau}_i\};\{\widehat{U}_i\}) = O(\delta(P)^{-1})$ {with high probability.}    
    Furthermore, there exists a bijection $\sigma:[K]\mapsto[K]$ such that 
    $\sum_{i=1}^n \mathbb{I}(\widehat{\tau}_i\neq \sigma(\tau_i)) = 0$
    with high probability. 
\end{theorem}

\begin{remark}
In this paper, we proposed and empirically evaluated a parametric bootstrap strategy for model selection, although we did not provide a formal proof of its validity. In recent years, there has been significant progress in the theoretical understanding of bootstrapping techniques for network data, with notable contributions from \cite{bhattacharyya2015subsampling,green2022bootstrapping,lundesubsampling,levin2025bootstrapping}. 
However, none of these existing results apply to the objective functions $Q_1$, $Q_2$, and $Q_3$ proposed in this paper.
As a result, new theoretical tools are required to establish bootstrap consistency for our proposed methodology. Developing such a foundation remains an important direction for future research.
\end{remark}

\section{Simulation study}
\label{sec:sim}
In this section, we assess the empirical performance of our proposed framework for community detection and model selection under the three blockmodels, as well as the nested blockmodel proposed by \citet{noroozi2021hierarchy}.
Wherever existing methods are available, we compare the proposed approach against state-of-the-art methods.
Note that there are no existing methods for testing the DCBM vs. the PABM.
The results demonstrate that our methodology achieves high accuracy both for label recovery and model selection consistency throughout.
In particular, the proposed workflow either outperforms or matches existing approaches across a range of network settings.

\subsection{Community detection}
We first analyze the performance of our proposed community detection methods for the three blockmodels. 
We refer to the community detection methods for SBM, DCBM, and PABM as $Q_1$, $Q_2$, and $Q_3$, respectively. 
The assessment metric of interest is the mislabeling rate, and
the results are presented in Tables \ref{tab:comdetsbm}, \ref{tab:comdetdcbm} and \ref{tab:comdetpabm}, respectively.
\label{subsec:comdet}
\subsubsection{Stochastic blockmodel}
We generated networks from the SBM with $K=3$ communities with 25\%, 25\% and 50\% nodes, respectively. 
The block probability matrix was
$\Omega\propto \begin{pmatrix}
    4 & 2 & 1\\
    2 & 4 & 1\\
    1 & 1 & 4
\end{pmatrix},$
$n$ is varied over $(1000, 2000, 3000)$ and the network density, $\delta$, is set to $0.05$. We simulated 100 networks for each combination of $(n,K)$. We compared the performance of $Q_1$ with spectral clustering using the Laplacian matrix (\textit{SC-L}) \citep{rohe2011spectral, sengupta2015spectral}, where the $K$-means algorithm is applied on the spectral embeddings of the Laplacian matrix $L=D^{-\frac{1}{2}}AD^{-\frac{1}{2}}$ instead of the adjacency matrix. 

\begin{table}[ht]
    \centering
    \footnotesize
    \begin{tabular}{|c|c|c|c|c|}
        \hline
        $n$ & $K$ & $\delta$ & $Q_1$ & \textit{SC-L} \\
        \hline
        1000 & 3 & 0.05 & 0.03 $\pm$ 0.006 & 0.02 $\pm$ 0.005 \\
        2000 & 3 & 0.05 & 0.00 $\pm$ 0.001 & 0.00 $\pm$ 0.001 \\
        3000 & 3 & 0.05 & 0.00 $\pm$ 0.000 & 0.00 $\pm$ 0.000 \\
        \hline
    \end{tabular}
    \caption{\small Average mislabeling error $\pm$ s.d. under SBM for $Q_1$ and \textit{SC-L}}
    \label{tab:comdetsbm}
\end{table}

The average mislabeling errors (proportion of mislabeled nodes) from applying $Q_1$ and \textit{SC-L} on the networks are reported in Table \ref{tab:comdetsbm}. We observe that as $n$ increases, $Q_1$ recovers the true communities perfectly, that is, the proportion of mislabeled nodes goes to zero. We also find that for all choices of $(n,K)$, there is no significant difference between $Q_1$ and \textit{SC-L} in terms of mislabeling errors. 

\subsubsection{Degree-corrected blockmodel}
We generated networks from the DCBM with a similar configuration as before. The community sizes and the block probability matrix remains the same, 
and the degree parameters $\{\theta_i\}$ are sampled independently from the $\mathrm{Beta}(1,5)$ distribution to emulate power-law type behavior in node degrees. 
The number of nodes, $n$, is varied over $(1000, 2000, 3000)$ and the network density, $\delta$, is set to $0.05$. 
We compared the results to regularized spectral clustering using the Laplacian matrix (\textit{RSC-L}), where the spectral embeddings of the Laplacian matrix are normalized before applying the $K$-means algorithm. 
The additional normalization step removes the effect of the multiplicative factor in the spectral embeddings for DCBM.

\begin{table}[ht]
    \centering
    \footnotesize
    \begin{tabular}{|c|c|c|c|c|c|c|}
        \hline
        $n$ & $K$ & $\delta$ & $Q_2$ & runtime & \textit{RSC-L} & runtime \\
        \hline
        1000 & 3 & 0.05 & 0.10 $\pm$ 0.010 & 0.029 & 0.08 $\pm$ 0.009 & 0.027 \\
        2000 & 3 & 0.05 & 0.05 $\pm$ 0.006 & 0.057 & 0.04 $\pm$ 0.005 & 0.066 \\
        3000 & 3 & 0.05 & 0.04 $\pm$ 0.004 & 0.088 & 0.03 $\pm$ 0.003 & 0.129 \\
        \hline
    \end{tabular}
    \caption{\small Average mislabeling error $\pm$ s.d. and runtime (seconds) under DCBM for $Q_2$ and \textit{RSC-L}}
    \label{tab:comdetdcbm}
\end{table}
From Table \ref{tab:comdetdcbm}, we observe that as $n$ increases, the proportion of mislabeled nodes for both $Q_2$ and \textit{RSC-L} decreases. Moreover, there is no significant difference between $Q_2$ and \textit{RSC-L} in terms of the mislabeling errors. In terms of runtime, we observe that the runtimes of $Q_2$ and $RSC$-$L$ are also of similar order.

\subsubsection{Popularity-adjusted blockmodel}

We generated networks from the PABM with $n$ nodes and $K$ equal-sized communities, varying $n$ over $(600,900,1500)$ and $K$ over $(2,3)$.
The node popularity matrix $\Lambda$ was chosen such that
$\Lambda=\begin{pmatrix}
    \mathbf{v}_{11} & \mathbf{v}_{12}\\
    \mathbf{v}_{21} & \mathbf{v}_{22}
 \end{pmatrix},
$ for $K=2$
and 
$\Lambda=\begin{pmatrix}
    \mathbf{v}_{11} & \mathbf{v}_{12} & \mathbf{v}_{13}\\
    \mathbf{v}_{21} & \mathbf{v}_{22} & \mathbf{v}_{23}\\
    \mathbf{v}_{31} & \mathbf{v}_{32} & \mathbf{v}_{33}
 \end{pmatrix}$
 for $K=3$,
where the vectors $\mathbf{v}_{k\ell}\in\real^{n_k},\ 1\leq k,\,\ell\leq K$. 
The elements of $\mathbf{v}_{kk}$ are generated independently from the $\mathrm{Beta}(2,1)$ distribution and for $k\neq \ell$, the elements of $\mathbf{v}_{k\ell}$ are generated independently from the $\mathrm{Beta}(1,2)$ distribution, imposing a homophilic community structure. 
We applied $Q_3$ to 100 networks simulated for each combination of $(n,K)$, and compared it to the Orthogonal Subspace Clustering algorithm (\textit{OSC}) proposed in \citet{koo2022popularity}.
\begin{table}[ht]
    \centering
    \footnotesize
    \begin{tabular}{|c|c|c|c|c|c|c|c|}
        \hline
        $n$ & $K$ & $\delta$ & $Q_3$ & runtime & \textit{OSC} & runtime\\
        \hline
        600 & 2 & 0.28 & 0.02 $\pm$ 0.007 & 0.032 & 0.02 $\pm$ 0.056 & 1.172\\
        900 & 2 & 0.28 & 0.01 $\pm$ 0.003 & 0.038 & 0.01 $\pm$ 0.042 & 3.592\\
        1500 & 2 & 0.28 & 0.01 $\pm$ 0.001 & 0.059 & 0.00 $\pm$ 0.001 & 16.808\\
        [.2cm]
        600 & 3 & 0.33 & 0.13 $\pm$ 0.057 & 1.052 & 0.09 $\pm$ 0.104 & 1.247\\
        900 & 3 & 0.33 & 0.02 $\pm$ 0.014 & 1.066 & 0.02 $\pm$ 0.060 & 3.690\\
        1500 & 3 & 0.33 & 0.01 $\pm$ 0.002 & 0.998 & 0.00 $\pm$ 0.001 & 16.976\\
        \hline
    \end{tabular}
    \caption{\small Average mislabeling error $\pm$ s.d. and runtime in seconds under PABM for $Q_3$ and \textit{OSC}}
    \label{tab:comdetpabm}
\end{table}

In Table \ref{tab:comdetpabm}, we see that under both settings, the average mislabeling error for both $Q_3$ and \textit{OSC} goes to zero as $n$ increases. 
We also observe that although the average error rates for the two methods are similar, the standard deviations for \textit{OSC} are slightly `higher' than $Q_3$ for most of the cases (except $n=1500, K=2$).
To understand this pattern, we looked into the individual error rates for both methods.
We found that for a few of the replications, \textit{OSC} returned very poor community estimates, thus resulting in a high error, while for the rest of the replications, it performed very well, often better than $Q_3$.
On the other hand, the performance of $Q_3$ is more consistent and is less affected by such bad samples.
From a methodological perspective, the two algorithms use the same matrix $\widehat{U}$ of the leading eigenvector of $A$ for clustering.
$Q_3$ attempts to cluster the rows of $\widehat{U}$ into subspaces via a $K$-means type algorithm, except that we have distinct \textit{projection matrices} instead of distinct \textit{mean vectors} corresponding to the clusters.
Whereas, \textit{OSC} attempts to cluster the rows of $\widehat{U}\widehat{U}^\top$ via spectral clustering (spectral decomposition + standard $K$-means).
Thus, apart from the accuracy discussion, $Q_3$ is also computationally less expensive than \textit{OSC} since, along with the spectral decomposition of $A$, \textit{OSC} involves the extra step of performing spectral decomposition of the matrix $\widehat{U}\widehat{U}^\top$.
In terms of runtime, we find that the proposed algorithm $Q_3$ is significantly faster than \textit{OSC}, confirming the scalability suggested by the computational complexity analysis in Remark \ref{rem:comp_complexity}.
Interestingly, for the simulation results with $K=3$, the runtimes for $Q_3$ do not increase with $n$, likely because the maximum number of iterations is reached before convergence.
But in general, we expect the runtimes to increase with $n$.

Our unified framework provides three objective functions for each of the three blockmodels under consideration, allowing for a natural way to do both community detection and model selection for them. 
As we have found in this subsection, the derived community detection methods are at par with some of the existing community detection methods we have for SBM, DCBM, and PABM. 
In the next subsection, we address the problem of model selection for blockmodels, where we believe the main contribution of our paper lies.

\subsection{Model selection}

In this subsection, we study the performance of our proposed two-step testing procedure for model selection. 
We refer to the tests corresponding to the testing problems in \eqref{test:sbmvsdcbm} (SBM vs. DCBM) and \eqref{test:dcbmvspabm} (DCBM vs. PABM) as $Q_1$ and $Q_2$ respectively.
The assessment metrics of interest are the Type-1 error rate and the power of the tests.

\subsubsection{SBM vs. DCBM}
We generated networks from the SBM and the DCBM with $n=600$ nodes and $K$ equal-sized communities. 
The block probability matrix is $\Omega\propto (1-\beta)\,I+\beta\,\mathbf{1}\mathbf{1}^\top$, where
$\beta$ is the ratio of the between-block probability and the within-block probability of an edge. The smaller the value of $\beta$ is, the easier it should be to detect the communities.
For the DCBM, the degree parameters $\{\theta_i\}$ are simulated from the power-law distribution with lower bound 1 and scaling parameter 5. 
We vary $K$ over $(3, 5)$, $\beta$ over $(0.2, 0.5)$, and the average degree of the network over $(15, 20, 40)$.
For each combination of $(n,\,K,\,\beta,\,\text{avg. degree})$, we simulated 100 networks and applied $Q_1$.
200 bootstrap samples are used 
to estimate the $p$-value of the test, and $H_0$ is rejected when the $p$-value falls below 0.05. 

\begin{table}[ht]
    \centering
    \footnotesize
    \begin{tabular}{|c|c|c|c|c|c|c|c|c|}
        \hline
        \multicolumn{9}{|c|}{True model: SBM}\\
        \hline
        $n$ & $K$ & $\beta$ & avg. degree & {$Q_1$} & \textit{EigMax} & \textit{EigMax-boot} & \textit{ECV-}$L_2$ & \textit{ECV-dev} \\
        \hline
        600 & 3 & 0.2 & 15 & 0.00 & 1.00 & 0.00 & 0.00 & 0.00 \\
        600 & 3 & 0.2 & 20 & 0.03 & 1.00 & 0.00 & 0.00 & 0.00 \\
        600 & 3 & 0.2 & 40 & 0.05 & 0.48 & 0.01 & 0.00 & 0.00 \\
        [.2cm]
        600 & 3 & 0.5 & 15 & 0.03 & 0.71 & 0.00 & 0.00 & 0.00 \\
        600 & 3 & 0.5 & 20 & 0.02 & 0.50 & 0.00 & 0.00 & 0.00 \\
        600 & 3 & 0.5 & 40 & 0.00 & 0.21 & 0.00 & 0.00 & 0.00 \\
        [.2cm]
        600 & 5 & 0.2 & 15 & 0.00 & 1.00 & 0.00 & 0.00 & 0.00 \\
        600 & 5 & 0.2 & 20 & 0.00 & 0.95 & 0.00 & 0.00 & 0.00 \\
        600 & 5 & 0.2 & 40 & 0.00 & 0.36 & 0.03 & 0.00 & 0.00 \\
        [.2cm]
        600 & 5 & 0.5 & 15 & 0.04 & 0.18 & 0.00 & 0.00 & 0.00 \\
        600 & 5 & 0.5 & 20 & 0.02 & 0.03 & 0.00 & 0.00 & 0.00 \\
        600 & 5 & 0.5 & 40 & 0.00 & 0.01 & 0.00 & 0.00 & 0.00 \\
        \hline
    \end{tabular}
    \caption{\small Proportion of times $H_0:A\sim\text{SBM}$ is rejected by $Q_1$ and \textit{EigMax}, and DCBM is selected by \textit{ECV}, when the true model is SBM.}
    \label{tab:sbmvsdcbm1}
\end{table}

\begin{table}[ht]
    \centering
    \footnotesize
    \begin{tabular}{|c|c|c|c|c|c|c|c|c|}
    \hline
        \multicolumn{9}{|c|}{True model: DCBM}\\
        \hline
        $n$ & $K$ & $\beta$ & avg. degree & {$Q_1$} & {\textit{EigMax}} & {\textit{EigMax-boot}} & {\textit{ECV-}$L_2$} & {\textit{ECV-dev}} \\
        \hline
        600 & 3 & 0.2 & 15 & 1.00 & 1.00 & 0.66 & 0.87 & 0.71 \\
        600 & 3 & 0.2 & 20 & 1.00 & 1.00 & 1.00 & 1.00 & 0.96 \\
        600 & 3 & 0.2 & 40 & 1.00 & 1.00 & 1.00 & 1.00 & 1.00 \\
        [.2cm]
        600 & 3 & 0.5 & 15 & 0.80 & 1.00 & 0.48 & 0.43 & 0.29 \\
        600 & 3 & 0.5 & 20 & 0.87 & 1.00 & 0.67 & 0.90 & 0.88 \\
        600 & 3 & 0.5 & 40 & 0.98 & 1.00 & 0.97 & 1.00 & 1.00 \\
        [.2cm]
        600 & 5 & 0.2 & 15 & 0.64 & 1.00 & 0.16 & 0.97 & 0.86 \\
        600 & 5 & 0.2 & 20 & 1.00 & 1.00 & 0.86 & 0.98 & 0.93 \\
        600 & 5 & 0.2 & 40 & 1.00 & 1.00 & 1.00 & 1.00 & 1.00 \\
        [.2cm]
        600 & 5 & 0.5 & 15 & 0.90 & 1.00 & 0.03 & 0.33 & 0.23 \\
        600 & 5 & 0.5 & 20 & 0.90 & 1.00 & 0.07 & 0.79 & 0.71 \\
        600 & 5 & 0.5 & 40 & 0.76 & 1.00 & 0.61 & 1.00 & 1.00 \\
        \hline
    \end{tabular}
    \caption{\small Proportion of times $H_0:A\sim\text{SBM}$ is rejected by $Q_1$ and and \textit{EigMax}, and DCBM is selected by \textit{ECV}, when the true model is DCBM.}
    \label{tab:sbmvsdcbm2}
\end{table}

We compared the performance of our method with two existing methods: (i) A goodness-of-fit test for SBMs proposed by \citet{lei2016goodness} and (ii) The edge cross-validation(\textit{ECV}) method for model selection proposed by \citet{li2020network}. 
\citet{lei2016goodness} proposed a test statistic based on the largest singular value of the residual adjacency matrix $\widetilde{A}_{ij}=(A_{ij} - \widehat{P}_{ij})/\sqrt{\widehat{P}_{ij}(1-\widehat{P}_{ij})}$, which has an asymptotic Tracy-Widom distribution when the true model is SBM.
Noting that the convergence might be slow, \citet{lei2016goodness} also proposed an alternative test statistic using bootstrap correction.
We consider both of the proposed test statistics for comparison, and call the tests \textit{EigMax} and \textit{EigMax-boot} respectively.
The ECV is a classification method which, given a set of candidate models (two, in our case), estimates a suitably chosen loss function using the adjacency matrix $A$ and selects the one with the minimum loss function value.
\citet{li2020network} analyzed two loss functions, a least-squared loss and a binomial deviance loss, and we call the corresponding methods \textit{ECV-$L_2$} and \textit{ECV-dev} respectively.
We fully acknowledge that the comparison our method to ECV is not an apples-to-apples comparison because our method is based on a testing framework, while the ECV is based on classification using a loss function minimization criterion.
Tables \ref{tab:sbmvsdcbm1} and \ref{tab:sbmvsdcbm2} compare the five model selection procedures when the data is generated from the SBM and the DCBM respectively.
For the testing procedures, $Q_1$, \textit{EigMax} and \textit{EigMax-boot}, we report the proportion of times $H_0:A\sim SBM$ is rejected. For \textit{ECV-$L_2$} and \textit{ECV-dev}, we report the proportion of times DCBM is selected as the true model.

In Table \ref{tab:sbmvsdcbm1}, we observe that when the true model is SBM, the size-estimates of $Q_1$ are always below the significance level of $0.05$. 
We find a similar result for \textit{EigMax-boot}, where the size-estimates also remain below the significance level of $0.05$. 
However, \textit{EigMax} performs very poorly, as we see that the size-estimates exceed the significance level in all cases.
Following \citet{lei2016goodness}, we attribute the poor performance of \textit{EigMax} to the slow convergence of the test statistic under the null.
Finally, the ECV always chose the correct model. 
This difference is purely because we calibrate the test for $\alpha=0.05$. 
The proposed test matches the target significance level, and if one wants a lower Type-1 error, this could be easily achieved by lowering the value of $\alpha$ and recalibrating the test.
Note that ECV, by design, does not offer this flexibility.

When the true model is DCBM and $\beta$ is 0.2, that is, there is high homophily, $Q_1$ has power equal to 1 in all cases. 
Even when $\beta$ is 0.5 (low homophily), although the problem becomes more challenging, the powers still remain above 0.8.
Moreover, $Q_1$ consistently has a higher power than \textit{EigMax-boot} in all cases.
\textit{EigMax}, on the other hand, has power equal to 1 across all cases, but its reliability is questionable.
As we found in Table \ref{tab:sbmvsdcbm1}, \textit{EigMax} may have size much larger than the significance level, that is, it may disproportionally reject $H_0:A\sim SBM$ even when the true model is SBM, with identical choices of parameters.
Finally, in comparison with ECV, we observe that the proportion of rejections in favor of $H_1:A\sim\text{DCBM}$ using $Q_1$ is mostly larger than the proportion of times DCBM is selected as the true model by ECV. 
However, as we said earlier, we should be cautious while interpreting the results from these two frameworks.

\subsubsection{DCBM vs. PABM}
We generated networks from the DCBM following the same configuration as before. Here, we kept $\beta$ fixed at 0.5, noting that it is the more difficult case to deal with. 
For PABM, we used the setting introduced in Section \ref{subsec:comdet} for community detection. Here, we fixed $n=900$ and scaled $P$ so that the network density $\delta$ varies over $(0.01, 0.02, 0.05, 0.1)$. 
For each scenario, we simulated 100 networks and applied $Q_2$. As before, we used 200 bootstrap samples to estimate the $p$-value of the test and rejected $H_0$ when it fell below 0.05. The results for DCBM and PABM are presented in Tables \ref{tab:dcbmvspabm1} and \ref{tab:dcbmvspabm2} respectively.

\begin{table}[ht]
    \centering
    \footnotesize
    \begin{tabular}{|c|c|c|c|c|}
        \hline
        $n$ & $K$ & $\beta$ & avg. degree &  $Q_2$ \\
        \hline
        600 & 3 & 0.5 & 15 & 0.02\\
        600 & 3 & 0.5 & 20 & 0.02\\
        600 & 3 & 0.5 & 40 & 0.01\\
              \hline
    \end{tabular}
    \begin{tabular}{|c|c|c|c|c|}
        \hline
        $n$ & $K$ & $\beta$ & avg. degree &  $Q_2$ \\
        \hline
           600 & 5 & 0.5 & 15 & 0.02\\
        600 & 5 & 0.5 & 20 & 0.00\\
        600 & 5 & 0.5 & 40 & 0.00\\
        \hline
    \end{tabular}
    \caption{\small Proportion of times $H_0:A\sim \text{DCBM}$ is rejected by $Q_2$ when the true model is DCBM}
    \label{tab:dcbmvspabm1}
\end{table}

\begin{table}[ht]
    \centering
    \footnotesize
    \begin{tabular}{|c|c|c|c|}
        \hline
        $n$ & $K$ & $\delta$ &  $Q_2$\\ 
        \hline
        900 & 2 & 0.01 & 0.91\\
        900 & 2 & 0.02 & 1.00\\
        900 & 2 & 0.05 & 1.00\\
        900 & 2 & 0.10 & 1.00\\
               \hline
    \end{tabular}
    \begin{tabular}{|c|c|c|c|}
        \hline
        $n$ & $K$ & $\delta$ &  $Q_2$\\ 
        \hline
        900 & 3 & 0.01 & 0.32\\
        900 & 3 & 0.02 & 1.00\\
        900 & 3 & 0.05 & 1.00\\
        900 & 3 & 0.10 & 1.00\\
        \hline
    \end{tabular}
    \caption{\small Proportion of times $H_0:A\sim\text{DCBM}$ is rejected by $Q_2$ when the true model is PABM}
    \label{tab:dcbmvspabm2}
\end{table}

We observe that when the true model is DCBM (Table \ref{tab:dcbmvspabm1}), the size of the test is always less than the significance level of $0.05$. 
When the true model is PABM (Table \ref{tab:dcbmvspabm2}), the power of the test is 1 except when the network density is small, i.e., the network is too sparse. 
If a network is too sparse, the latent position estimation as well as the community estimation problem becomes harder, which possibly leads to a small power in this case.

\subsubsection{Nested stochastic blockmodel}
Next, we applied $Q_2$ on networks generated from the nested stochastic blockmodel (NBM), proposed by \citet{noroozi2021hierarchy}. 
Observing the significant jump in the parameters from DCBM to PABM, the NBM was proposed as a bridge between the DCBM and PABM in the hierarchy of blockmodels. 
The NBM has $K$ communities and $L$ meta-communities, where each meta-community is composed of members from exactly one or more of the $K$ communities, that is, $L\leq K$. 
When $L=1$, the NBM reduces to the DCBM, and for $L=K$, the NBM becomes a PABM.
To generate networks from the NBM, we used the same simulation setting as the one considered in Section 7.1 of \cite{noroozi2021estimation}. We refer the readers to \cite{noroozi2021hierarchy} for details on generating the model parameters.
Along with $n, K, $ and $L$, there is an additional factor $\omega$ which captures the homophily in the network, such that as $\omega$ increases, the community estimation and, subsequently, the model selection task become harder. 
\begin{table}[ht]
    \centering
    \footnotesize
    \begin{tabular}{|c|c|c|c|Hc|}
        \hline
        $n$ & $K$ & $L$ & $\omega$ & $\delta$ &  $Q_2$\\ 
        \hline
        900 & 6 & 1 & 0.6 & 0.12 & 0.03\\ 
        900 & 6 & 1 & 0.8 & 0.15 & 0.18\\ 
        900 & 6 & 2 & 0.6 & 0.12 & 0.99\\ 
        900 & 6 & 2 & 0.8 & 0.15 & 1.00\\ 
        900 & 6 & 3 & 0.6 & 0.12 & 1.00\\ 
        900 & 6 & 3 & 0.8 & 0.15 & 1.00\\ 
        900 & 6 & 6 & 0.6 & 0.12 & 1.00\\ 
        900 & 6 & 6 & 0.8 & 0.15 & 1.00\\ 
        \hline
    \end{tabular}
    \begin{tabular}{|c|c|c|c|Hc|}
        \hline
        $n$ & $K$ & $L$ & $\omega$ & $\delta$ &  $Q_2$\\ 
        \hline
        1260 & 6 & 1 & 0.6 & 0.12 & 0.01\\ 
        1260 & 6 & 1 & 0.8 & 0.15 & 0.13\\ 
        1260 & 6 & 2 & 0.6 & 0.12 & 0.96\\ 
        1260 & 6 & 2 & 0.8 & 0.15 & 1.00\\ 
        1260 & 6 & 3 & 0.6 & 0.12 & 1.00\\ 
        1260 & 6 & 3 & 0.8 & 0.15 & 1.00\\ 
        1260 & 6 & 6 & 0.6 & 0.12 & 1.00\\ 
        1260 & 6 & 6 & 0.8 & 0.15 & 1.00\\ 
        \hline
    \end{tabular}
    \caption{\small Proportion of times $H_0:\text{DCBM}$ is rejected by $Q_2$ when the true model is NBM ($L=1$ corresponds to DCBM, and $L=K$ corresponds to PABM)}
    \label{tab:dcbmvspabm3}
\end{table}

In Table \ref{tab:dcbmvspabm3}, we observe that when $L=1$, that is, the true model is DCBM, the size of the test is below the significance level of 0.05 for $\omega=0.6$, that is, when there is high homophily. For $\omega=0.8$ (low homophily), the power is 0.18 and 0.13 when $n$ is $900$ and $1260$, respectively, which is larger than 0.05. 
Next, we observe that as $L$ increases, the model becomes more and more complex than the DCBM, and it gradually becomes easier for the test to correctly reject $H_0$. When $L=2$, the power of the test is high (above 0.96 for all values of $n$, $K$, and $\omega$), although still below 1.
When $L=3$ and $6$, the power of the test is 1, that is, the test is always rejected.

\subsubsection{Model selection for disassortative networks}

Here, we implemented the full model selection pipeline for dissortative networks from the SBM, the DCBM, and the PABM with $n=900$ nodes, $K=3$ equal-sized communities, and two values of density, $\delta = 0.05$ and $\delta = 0.1$.
For the SBM and the DCBM, we used 
$$\Omega \propto \begin{pmatrix}
    1/3 & 2/3 & 2/3\\
    2/3 & 1/3 & 2/3\\
    2/3 & 2/3 & 1/3
\end{pmatrix}.$$
The DCBM degree parameters were generated from the power-law distribution with lower bound 1 and scaling parameter 5.
For the PABM, we introduced dissortativity into the setting from Section \ref{subsec:comdet} by generating the elements of $\mathbf{v}_{kk}$ and $\mathbf{v}_{k\ell}$ (for $k\neq \ell$) from the $\mathrm{Beta}(1,2)$ and $\mathrm{Beta}(2,1)$ distributions, respectively.
\begin{table}[h]
    \centering
    \footnotesize
    \begin{tabular}{|c|c|c|c|c|c|c|}
        \hline
        \multicolumn{3}{|c|}{Attributes} & & \multicolumn{3}{c|}{Selected model} \\
        \hline
        $n$ & $K$ & $\delta$ & True model  & SBM & DCBM & PABM \\
        \hline
        900 & 3 & 0.05 & SBM & 1.00 & - & - \\
        900 & 3 & 0.1 & SBM & 1.00 & - & - \\
        900 & 3 & 0.05 & DCBM & 0.02 & 0.98 & - \\
        900 & 3 & 0.1 & DCBM & - & 1.00 & - \\
        900 & 3 & 0.05 & PABM & - & - & 1.00 \\
        900 & 3 & 0.1 & PABM & - & - & 1.00\\
        \hline
    \end{tabular}
    \caption{Model selection results for disassortative networks using Algorithm \ref{algo1}.}
    \label{tab:dissortative}
\end{table}

We generated 100 networks from each model and implemented Algorithm \ref{algo1}. 
Table \ref{tab:dissortative} reports the proportion of times SBM, DCBM, and PABM are selected.
For networks generated from SBM and PABM, the model is correctly found in all replications.
For networks generated from DCBM, the correct model is selected in 98 out of 100 replications in the sparser case ($\delta=0.05$) and in all replications in the denser case ($\delta=0.1$).
Thus, the proposed method performs accurate model selection for disassortative networks.

\section{Application to real-world datasets}
\label{sec:data}
We applied our model selection procedure to five well-studied real-world networks that are known to have a community structure.
For each network, we
implemented the pipeline in Figure \ref{fig:schema} to determine which of the three blockmodels would best describe the network. Below, we briefly describe each network and present the results of our analysis.

\textbf{Karate club}: 
The Karate club network represents social ties among $n=34$ members of a university Karate club  \citep{zachary1977information}. A conflict caused the club to split into $K=2$ factions, making it a classic example of community structure \citep{girvan2002community}.

\textbf{Dolphin}: The Dolphin network represents social links between $n=62$ bottlenose dolphins in Doubtful Sound, New Zealand, with $K=2$ communities \citep{lusseau2003emergent}.

\textbf{British MP}: This network represents retweets between $n=329$ Members of Parliament (MPs) from $K=2$ communities, corresponding to the Conservative and Labour Parties, the two largest political parties in the United Kingdom \citep{greene2013producing}.

\textbf{Political blogs}: The political blogs network consists of hyperlinks between $n=1222$ U.S. political blogs two months before the 2004 Presidential election \citep{adamic2005political}. The blogs are labeled as either liberal or conservative, representing $K=2$ communities \citep{karrer2011stochastic, amini2013, jin2015fast}.

\textbf{DBLP}: The DBLP network consists of $n=4057$ researchers from $K=4$ research areas:
{database}, {data mining}, {information retrieval}, and {artificial intelligence} \citep{gao2009graph, ji2010graph}.
Two researchers are connected if they published at the same conference \citep{yanchenko2024generalized,bhadra2025scalable,chakrabarty2025subsampling}. 
\begin{table}[ht]
    \centering
    \footnotesize
    \begin{tabular}{|c|r|c|r|c|c|c|c|}
    \hline
        Network &  \multicolumn{2}{c|}{Attributes} & \multicolumn{2}{c|}{$H_0:$ SBM vs. $H_1:$ DCBM} & \multicolumn{2}{c|}{$H_0:$ DCBM vs. $H_1:$ PABM}\\
        \hline
        & $n$ & $K$  & $H_0$ rejected? & $p$-value & $H_0$ rejected? & $p$-value\\
        \hline
        Karate club & 34 & 2  & \checkmark & 0.03 & \ding{55} & 0.29\\
        Dolphin & 62 & 2  & \checkmark & 0.01 & \ding{55} & 0.75\\
        British MP & 329 & 2  & \checkmark & 0.00 & \checkmark & 0.00\\
        Political blogs & 1222 & 2  & \checkmark & 0.00 & \checkmark & 0.00\\
        DBLP & 4057 & 4  & \checkmark & 0.00 & \checkmark & 0.00\\
        \hline
    \end{tabular}
    \caption{\small Model selection results from five real-world networks}
    \label{tab:data}
\end{table}

From Table  \ref{tab:data}, we observe that the SBM vs. DCBM test is rejected for all networks.
This is not surprising, because the SBM is a simplistic model that does not adequately capture the real-world characteristics of networks.
Next, the DCBM vs. PABM test is not rejected for the Karate Club and Dolphin networks. 
Therefore, the conclusion is that while the SBM is not good enough to model these two networks, the DCBM is able to explain the community structure very well. 
The DCBM vs. PABM test is rejected for the British MP network, the political blogs network, and the DBLP network, implying that these three networks can not be accurately represented by the DCBM.

\section{Discussion}
\label{sec:conc}
In this paper, we presented a unified framework to perform community detection and model selection by utilizing the nested structure of the blockmodels: SBM, DCBM, and PABM.
Through a detailed simulation study and real-world applications, we showed that given a network generated from one of these models, the proposed method is able to select the correct model as well as estimate the true community assignments accurately.
We also derived the theoretical properties of the proposed method under the three models. 

One important direction of future research will be to develop a formal proof of the parametric bootstrap strategy we used to estimate the $p$-value of our tests for model selection. 
Also, the proposed workflow does not include some notable variants of blockmodels, such as the mixed-membership stochastic blockmodel (MMSBM \cite{airoldi2009mixed}) and the degree-corrected mixed-membership model (DCMM, \cite{jin2023mixed}).
Another important direction of future research would be to extend the proposed framework to these models.

\bibliographystyle{chicagoa}

\bibliography{Bibliography-MM-MC,ref}


\newpage
\setcounter{page}{1}

  \setcounter{section}{0} 
  \setcounter{equation}{0}
  \renewcommand{\theequation}{A\arabic{equation}}
  \renewcommand{\thesection}{A\arabic{section}}

\begin{center}
		{\Large \bf Supplementary material for ``A Unified Framework for Community Detection and Model Selection in Blockmodels''}
	\end{center}


\section{Computational complexity comparisons}


An important feature of our unified framework is that model selection does not require the computation of an additional test statistic beyond the loss function minimization used for community detection.
Existing two-step procedures, in contrast, incur the additional cost of computing a test statistic.
A natural question is, does this extra cost  of computing the  test statistic matter from a practical implementation perspective?
In other words, is this extra cost high enough to make the proposed approach substantially more efficient than the two-step procedure? 

As an example, consider testing between the SBM and the DCBM using the \textit{EigMax} method proposed by \citet{lei2016goodness}. The computation of the \textit{EigMax} statistic requires forming the residual adjacency matrix,
\[
\widetilde{A}_{ij}=\frac{A_{ij}-\widehat{P}_{ij}}{\widehat{P}_{ij}(1-\widehat{P}_{ij})},
\]
which involves $\Theta(n^2)$ operations even if the observed adjacency matrix $A$ is sparse. By contrast, spectral clustering has computational cost 
\[
O\big(\mathrm{nnz}(A)\,K + nK^2\big),
\]
where $\mathrm{nnz}(A)$ denotes the number of non-zero entries in $A$. For sparse networks with maximum degree bounded by $n\rho_n$ where $\rho_n \to 0$, we have $\mathrm{nnz}(A)=O(n^2\rho_n)$ with high probability. In such cases, the cost of computing the \textit{EigMax} statistic can dominate the overall runtime, while in our approach the relevant quantity for model comparison is obtained directly as a by-product of loss function minimization.

To illustrate this effect, we conducted a runtime comparison with SBM networks of sizes $n=1000,2000,\ldots,10000$ with $K=5$ equal-sized communities, block probability matrix $\Omega\propto 0.8I + 0.2\mathbf{1}\mathbf{1}^{\top}$, and expected density $\delta=0.1$. Table~\ref{tab:timecomp} reports the average runtimes (over 100 replications) for loss function minimization ($Q_1$) and computation of the \textit{EigMax} statistic. Note that for \textit{EigMax}, we only include the cost of constructing $\widetilde{A}$ and computing its largest singular value, excluding the preliminary spectral clustering step needed to estimate $\widehat{P}$.

\begin{table}[ht]
    \centering
    \resizebox{\textwidth}{!}{%
    \begin{tabular}{|c|c|c|c|c|c|c|c|c|c|c|}
    \hline
        n & 1000 & 2000 & 3000 & 4000 & 5000 & 6000 & 7000 & 8000 & 9000 & 10000 \\
        \hline
        $Q_1$ & 0.016 & 0.036 & 0.068 & 0.113 & 0.182 & 0.266 & 0.362 & 0.479 & 0.607 & 0.746 \\
        \textit{EigMax} & 0.249 & 1.727 & 4.069 & 8.805 & 13.174 & 18.559 & 29.637 & 38.262 & 47.072 & 64.859 \\
        \hline
    \end{tabular}%
    }
    \caption{Average runtimes (in seconds) of loss function minimization ($Q_1$) and test statistic computation for \textit{EigMax}. Results are averaged over 100 replications.}
    \label{tab:timecomp}
\end{table}

The results show that the runtime for test statistic computation grows much faster with network size (more than $85$ times for $n=10000$) than that for loss function minimization. This demonstrates that in sparse regimes, our unifying framework not only provides a principled way of performing model selection but can also yield substantial computational savings.

We would like to add that, in general, there could be alternative test statistics such that extra expense for computing a new test statistic is negligible compared to the loss function.
In that case, the proposed framework may not lead to a substantial improvement in computational efficiency from the perspective of
practical implementation.

\section{Proof of Theorem \ref{thm:q1_sbm}}
\begin{proof}
We prove Theorem \ref{thm:q1_sbm} for a global minimizer $\{\widehat\tau_i\}$ of $Q_1(\{t_i\};\{\widehat{U}_i\})$.
We first show that $Q_1(\{\widehat{\tau}_i\};\{\widehat{U}_i\})=O(\delta(P)^{-1})$ with high probability, and this result, combined with the concentration of the latent positions $\{\widehat{U}_i\}$'s, ensures the exact recovery of $\{\widehat{\tau}_i\}$.
For a $(1+\epsilon)$-approximation solution $\{\widehat{\tau}_i\}$, the proof can be carried out in the same manner, since $Q_1(\{\widehat{\tau}_i\};\{\widehat{U}_i\})=O(\delta(P)^{-1})$ still holds with high probability, by definition in Eq. \eqref{main.apx.sol.1}.

For $k\in[K]$, define
\begin{gather*}
\CC^{(\tau,\widehat{U})}_k = \frac{1}{n_k} \sum_{i \colon \tau_i = k} \widehat{U}_i,
\end{gather*}
where $n_k = |\{i \colon \tau_i = k\}|$, i.e., 
$\{\CC^{(\tau,\widehat{U})}_{k}\}$ are the cluster centroids for the
$\{\widehat{U}_i\}$ as given by $\{\tau_i\}$. 
Next recall Eq.~(\ref{eq:def_U_H}). Then by Lemma \ref{lem:estimation_U}, we have
\begin{equation}
  \label{eq:bdd_sbmq1}
  \begin{split}
  Q_1(\{\widehat{\tau}_i\};\{\widehat{U}_i\}) &\leq Q_1(\{\tau_i\};\{\widehat{U}_i\}) 
  = \sum_{k=1}^K \sum_{i \colon \tau_i = k} \|\widehat{U}_i - \CC^{(\tau,\widehat{U})}_{k} \|^2 \\ & \leq
  \sum_{k=1}^K \sum_{i\colon \tau_i = k} \|\widehat{U}_i - U_i\|^2 =\|\widehat{U} - U\|_F^2 = O(\delta(P)^{-1})
    \end{split}
\end{equation}
with high probability, under Assumtions \textbf{A1} and \textbf{A2}. 

The exact recovery of $\{\widehat{\tau}_i\}$ can be shown using the same argument as that for the proof of Theorem~6 in \cite{lyzinski2014perfect}. More specifically, let 
\[ r_* = \frac{1}{4} \min_{\tau_i \not = \tau_j} \|U_i - U_{j}\| = \frac{1}{4} \min_{\tau_i \not = \tau_j} \bigl(\|U_i\|^2 + \|U_j\|^2\bigr)^{\frac{1}{2}} \geq 
(8\,n_{\max})^{-\frac{1}{2}}.
\]
Note that in the above derivations we have used the fact that $K$ has orthonormal columns so that $H_k^{\top} H_{\ell} = 0$ whenever $k \not = \ell$ together with the form for $U$ in Eq.~\eqref{eq:def_U_H}.
Next let $\mathcal{B}_1, \dots, \mathcal{B}_K$ be $K$ balls of radii $r_*$ centered around the distinct rows of $U$; these balls are disjoint due to the choice of $r_*$.
Define
$$\CC^{(\widehat\tau,\widehat{U})}_{k} = \frac{1}{\widehat{n}_{k}} \sum_{i \colon \widehat{\tau}_i = k} \widehat{U}_i, \quad \text{for $1 \leq k \leq K$}, \\
$$ where $\widehat{n}_k = |\{i \colon \widehat\tau_i = k\}|$ and let $\widehat{C}$  be the $n \times K$ matrix with rows
$\widehat{C}_i = \CC^{(\widehat\tau,\widehat{U})}_{\widehat{\tau}_i}$. 

Now condition on the high-probability event in Eq.~\eqref{eq:bdd_sbmq1}. 
Next, suppose there exists a $k \in [K]$
such that $\mathcal{B}_{k}$ does not contain any row of $\widehat{C}$. 
Then $\|\widehat{C} - U\|_{F} \geq r_* \sqrt{n_{\min}}$ where $n_{\min} = \min\limits_{k\in[K]} n_k$ is the size of the smallest community. We thus have
\begin{align*}
    (Q_1(\{\widehat{\tau}_i\}; \{\widehat{U}_i\}))^{\frac{1}{2}} =& \|\widehat{C} - \widehat{U}\|_F \geq \|\widehat{C} - U\|_F - \|\widehat{U} - U\|_F \\
    \geq&  r_* \sqrt{n_{\min}} - O(\delta(P)^{-\frac{1}{2}}) = \omega((\delta(P))^{-\frac{1}{2}}),
\end{align*}
a contradiction, where the final inequality is due to the fact that $n_{\min} \asymp n_{\max}$ and so $r_* \sqrt{n_{\min}} \asymp 1$.
Therefore, by the pigeonhole principle, each ball $\mathcal{B}_k$
contains precisely one unique row of $\widehat{C}$. 

Choose an arbitrary pair $i \not =j$. Recall that, for the $K$-means clustering criterion,
each point $x$ is assign to the closest cluster centroid. 
Now if $\widehat{C}_i  = \widehat{C}_j$, then
$\widehat{U}_i$ and $\widehat{U}_j$ are assigned to the same cluster and hence they both belong
to $\mathcal{B}_{k}$ for some $k$, i.e., $\|\widehat{U}_i - \widehat{U}_j\| \leq 2\,r_*$.
Lemma~\ref{lem:estimation_U} then implies 
\[ \|U_i - U_j\| \leq 2\,\|\widehat{U} - U\|_{2 \to \infty} + 2\,r_* = O\left(\frac{\log^{\frac{1}{2}}{n}}{n^{\frac{1}{2}} \delta(P)^{\frac{1}{2}}}\right) + 
2\,r_* \leq 3\,r_* \]
with high probability and hence $U_i = U_j$ as the smallest gap between any two distinct rows of $U$ is at least $4\,r_*$.
Conversely, suppose $U_i = U_j$. Then, as $\|\widehat{U} - U\|_{2 \to \infty} \leq r_*$ with high probability, both $\widehat{U}_i$ and $\widehat{U}_j$ belongs to the same $\mathcal{B}_{\tau_i}$, and since $\mathcal{B}_{\tau_i}$ contains a unique row of $\widehat{C}$, both $\widehat{U}_i$ and $\widehat{U}_j$ will be assigned to the same cluster so that
$\widehat{C}_i = \widehat{C}_j$. 

In summary
$U_i = U_j$ if and only if $\widehat{C}_i = \widehat{C}_j$. We thus have $\tau_i = \tau_j$ if and only if
$\widehat{\tau}_i = \widehat{\tau}_j$ and hence
there exists a permutation $\sigma \colon [K] \mapsto [K]$ such that $\tau_{i} = \sigma(\widehat{\tau}_i)$ for all $i$.
\end{proof}

\section{Proof of Theorem \ref{thm:q1_dcbm}}

\begin{proof} First suppose that
\begin{equation}
    \label{q1txcond}
    Q_1(\{t_i\};\{{U}_i\})\gg 1/\delta(P) \ \text{for any }\{t_i\}.
\end{equation}
For $k\in[K]$, define
\begin{gather*}
\CC^{(t,U)}_k = \frac{1}{|\{i\colon t_i=k\}|} \sum_{i \colon t_i = k} U_i,\quad \CC^{(t,\widehat{U})}_k = \frac{1}{|\{i\colon t_i=k\}|} \sum_{i \colon t_i = k} \widehat{U}_i.
\end{gather*}
Then 
\begin{equation*}
  \begin{split}
  Q_1(\{t_i\};\{U_i\}) &= \sum_{i=1}^n \|U_i - \CC^{(t,U)}_{t_i}\|^2 \leq \sum_{i=1}^n \|U_i - \CC^{(t,\widehat{U})}_{t_i}\|^2 \\
  &\leq 2\sum_{i=1}^n \|U_i - \widehat{U}_i\|^2 + 2\sum_{i=1}^n
  \|\widehat{U}_i - \CC^{(t,\widehat{U})}_{t_i}\|^2\\
  &=  2\,\|\widehat{U} - U\|_{F}^2 +2\,Q_1(\{t_i\},\{\widehat{U}_i\})
         = O(\delta(P)^{-1}) + 2\,Q_1(\{t_i\}, \{\widehat{U}_i\}),
\end{split}
\end{equation*}
with high probability under Assumptions \textbf{A1} and \textbf{A2}, by Lemma \ref{lem:estimation_U}. 
Hence, $$Q_1(\{t_i\};\{\widehat{U}_i\})\gg 1/(\delta(P))$$
with high probability.

We now show Eq.~\eqref{q1txcond}.
Suppose to the contrary, there exists a finite constant $C$ such that
$Q_1(\{t_i\};\{{U}_i\}) \leq C/\delta(P)$. 
Note that by \eqref{eq:def_U_H_dcbm}, for any pair of nodes $i$ and $j$ with $\tau_i\neq \tau_{j}$, we have
\begin{equation}
  \label{eq:min_dist_U}
\umod{U_i-U_j}^2\,=\, (\widetilde{\theta}_i^2 + \widetilde{\theta}_{j}^2)\,\geq\, \frac{2\,\theta_{\min}^2}{n_{\max}}.
\end{equation}
Denote $r_* = \theta_{\min}^2/(3\,n_{\max})$ and
let $\mathcal{T}$ be the set of nodes for which 
$\|{U}_i - \CC^{(t,U)}_{t_i}\|^2\leq r_*$; note that 
$$|\mathcal{T}^{c}| \leq \frac{3\,C\,n_{\max}}{\theta_{\min}^2\,\delta(P)}.$$  
Then for any pair $(i,j) \in \mathcal{T} \times \mathcal{T}$ with $\tau_i \not = \tau_j$, we have by Eq.~\eqref{eq:min_dist_U} that
$\CC^{(t,U)}_{t_i} \not = \CC^{(t,U)}_{t_j}$. 
As there are exactly $K$ centroids, there has to be a bijection $\sigma:[K]\mapsto[K]$ such that for
all $i\in\mathcal{T}$, we have $\|{U}_i - \CC^{(t,U)}_{t_i}\|^2 =\|{U}_i - \CC^{(t,U)}_{\sigma(\tau_i)}\|^2$.

Let $n_{k,\mathcal{T}} = |\{i \in \mathcal{T} \colon \tau_i = k\}|$. 
We then have
\begin{equation}
  \begin{split}
    Q_1(\{t_i\},;\{{U}_i\})=\sum_{i=1}^n \|{U}_i - \CC^{(t,U)}_{t_i}\|^2 &\,\geq\, \sum_{k=1}^K \sum_{i\in \mathcal{T}\colon\tau_i=k} \|{U}_i - \CC^{(t,U)}_{\sigma(k)}\|^2\\
    &\,\geq\, \sum_{k=1}^K \sum_{i\in \mathcal{T}\colon\tau_i=k} \Bigl\|{U}_i - \frac{1}{n_{k,\mathcal{T}}} \sum_{j\in \mathcal{T},\tau_j=k} U_j \Bigr\|^2 \\
    &\,=\, \sum_{k=1}^K \sum_{i\in \mathcal{T}\colon\tau_i=k} \Bigl\|\widetilde{\theta}_i H_k - \frac{1}{n_{k,\mathcal{T}}} \sum_{j\in \mathcal{T},\tau_j=k}\widetilde{\theta}_j H_k\Bigr\|^2\\
    &\,=\, \sum_{k=1}^K \sum_{i\in \mathcal{T}\colon\tau_i=k} \Bigl\|\frac{\theta_i}{\umod{\phi}_k} H_k - \frac{1}{n_{k,\mathcal{T}}} \sum_{j\in \mathcal{T},\tau_j=k}\frac{\theta_j}{\umod{\phi}_k} H_k\Bigr\|^2\\
    &\,\geq\, \frac{1}{n_{\max}}\sum_{k=1}^K \sum_{i\in \mathcal{T}\colon\tau_i=k}(\theta_i - \bar{\theta}_{k,\mathcal{T}})^2,
\end{split}
\label{q1tx}
\end{equation}
where $\bar{\theta}_{k,\mathcal{T}}=\frac{1}{n_{k,\mathcal{T}}} \sum_{j\in \mathcal{T}\colon\tau_j=k}\theta_j$ and the second inequality in the above display
follows from the fact that $\frac{1}{n_{k, \mathcal{T}}} \sum_{j \in \mathcal{T}, \tau_j = k} U_j$ is the minimizer of $\sum_{i \in \mathcal{T} \colon \tau_i = k} \|U_i - \xi\|^2$ over all $\xi \in \mathbb{R}^K$. 

Now,
\begin{align}
    \sum_{k=1}^K \sum_{i\in \mathcal{T}\colon\tau_i=k}(\theta_i - \bar{\theta}_{k,\mathcal{T}})^2 
    &= \sum_{k=1}^K \Bigl(\sum_{i\colon\tau_i=k}(\theta_i - \bar{\theta}_{k,\mathcal{T}})^2 -  \sum_{i\in\mathcal{T}^c\colon\tau_i=k}(\theta_i - \bar{\theta}_{k,\mathcal{T}})^2 \Bigr) \nonumber\\
    &\geq \sum_{k=1}^K \sum_{i\colon\tau_i=k}(\theta_i - \bar{\theta}_{k,\mathcal{T}})^2 - |\mathcal{T}^{c}| \geq \sum_{k=1}^K
    \sum_{i\colon\tau_i=k}(\theta_i - \bar{\theta}_{k})^2 - |\mathcal{T}^{c}|.
    \label{thetasd2}
\end{align}
From \eqref{q1tx} and \eqref{thetasd2}, we have
\begin{equation}
  \begin{split}
    Q_1(\{t_i\};\{{U}_i\}) &\geq \frac{1}{n_{\max}}\sum_{k=1}^K\sum_{i\colon\tau_i=k}(\theta_i - \bar{\theta}_{k})^2 - \frac{|\mathcal{T}^{c}|}{n_{\max}} \\ & \geq \frac{1}{n_{\max}}\sum_{k=1}^K\sum_{i\colon\tau_i=k}(\theta_i - \bar{\theta}_{k})^2 - \frac{3}{\theta_{\min}^2 \delta(P)}.
    \end{split}
    \label{eq:q1_dcbm1}
\end{equation}
Leveraging Eq.~\eqref{thetavarcond} and noting that $\theta_{\max}/\theta_{\min} = O(1)$, we obtain
\begin{equation*}
    Q_1(\{t_i\};\{{U}_i\})\gg \frac{1}{\delta(P)},
\end{equation*}
which is a contradiction. 
\end{proof}

\section{Proof of Theorem \ref{thm:q2_dcbm}}
\begin{proof} We prove Theorem \ref{thm:q2_dcbm} for a global minimizer $\{\widehat\tau_i\}$ of $Q_2(\{t_i\};\{\widehat{U}_i\})$.
We first show that $Q_2(\{\widehat{\tau}_i\};\{\widehat{U}_i\})=O(\delta(P)^{-1})$ with high probability, and this result, combined with the concentration of the latent positions $\{\widehat{U}_i\}$'s, ensures the exact recovery of $\{\widehat{\tau}_i\}$.
For a $(1+\epsilon)$-approximation solution $\{\widehat{\tau}_i\}$, the proof can be carried out in the same manner, since $Q_2(\{\widehat{\tau}_i\};\{\widehat{U}_i\})=O(\delta(P)^{-1})$ still holds with high probability, by definition in Eq. \eqref{main.apx.sol.2}.

Recall that the objective function $Q_2$ is defined as
$$ Q_2(\{t_i\};\{\widehat{U}_i\})=\sum_{k=1}^K \sum_{i\colon t_i=k} \|(I - {\Pi}_{k}^{(t,\widehat{U})})\, \widehat{U}_i\|^2.$$ 
As $\{\widehat{\tau}_i\}$ is the global minimizer of $Q_2$,
$Q_2(\{\widehat{\tau}_i\};\{\widehat{U}_i\})\leq Q_2(\{{\tau}_i\};\{\widehat{U}_i\}).$
We now show that $Q_2(\{\tau_i\};\{\widehat{U}_i\})$
is small. 
Let us define the true projection matrices corresponding to the $K$ communities as,
$$\Gamma_k=H_kH_k^\top,\ 1\leq k \leq K.$$
We then have
\begin{align*}
  \,Q_2(\{\tau_i\};\{\widehat{U}_i\})&=\sum_{k=1}^K \sum_{i\colon{\tau}_i={k}} \|(I - {\Pi}_{k}^{(\tau,\widehat{U})})\,
    \widehat{U}_i\|^2\\
    &\leq  2\,\sum_{k=1} \sum_{i\colon{\tau}_i={k}} \|(I - \Pi_{k}^{(\tau,\widehat{U})})\, (\widehat{U}_i-U_i)\|^2+2\sum_{k=1}^K \sum_{i\colon{\tau}_i={k}} \|(\Gamma_{k} - \Pi_{k}^{(\tau,\widehat{U})})\, U_i\|^2\\
    &\leq \, 2\,\|\widehat{U}-U\|_F^2+2\sum_{k=1}^K \sum_{i\colon{\tau}_i=k}\|U_i\|^2\, \|\Pi_{k}^{(\tau,\widehat{U})}-\Gamma_{k}\|^2\\
    &\leq \, 2\,\|\widehat{U}-U\|_F^2+2\sum_{k=1}^K \sum_{i\colon{\tau}_i=k}\widetilde{\theta}_i^2\, \|\Pi_{k}^{(\tau,\widehat{U})}-\Gamma_{k}\|^2\ \text{(recall $\|H_k\|=1$)}\\
    &=\,2\,\|\widehat{U}-U\|_F^2+2\sum_{k=1}^K \|\Pi_{k}^{(\tau,\widehat{U})}-\Gamma_{k}\|^2.
\end{align*}
Fix some $k \in [K]$ and let $M_{k}$ be a matrix whose columns are all $U_i$'s
for which $\tau_i= k$. 
Similarly, let $\widehat{M}_{k}$ be the matrix whose columns are all $\widehat{U}_i$'s for which $\tau_i= k$.
Then $\Gamma_{k}$ and $\Pi_{k}^{(\tau,\widehat{U})}$ correspond to the orthorgonal projections onto the leading left singular vectors of $M_{k}$ and $\widehat{M}_{k}$ respectively. 
As $M_{k}$ has rank $1$, we have by the Wedin $\sin$-$\Theta$ theorem (see page $262$ of \cite{stewart_sun})
that 
$$\|\Pi_{k}^{(\tau,\widehat{U})}-\Gamma_{k} \| \leq \frac{\|\widehat{M}_{k}- M_{k} \|}{\|M_{k}\|}
= \|\widehat{M}_{k} - M_{k}\|,$$
where the final equality follows from the fact that $U_i = \widetilde{\theta}_i H_{\tau_i}$ and $\sum_{i \colon \tau_i = k} \widetilde{\theta}_i^2 = 1$ for all $k \in [K]$ (recall $\|H_k\|=1$). 
We therefore have
$$\sum_{k=1}^K \|\Pi_{k}^{(\tau,\widehat{U})}-\Gamma_{k}\|^2 \leq \sum_{k=1}^K \|\widehat{M}_{k} - M_{k}\|_{F}^2 = \|\widehat{U} - U\|_{F}^2.$$
Combining the above bounds together with Lemma \ref{lem:estimation_U}, we obtain that under Assumptions \textbf{A1} and \textbf{A2},
$$Q_2(\{\tau_i\}; \{\widehat{U}_i\}) \leq 4 \|\widehat{U} - U\|_{F}^2 = O(\delta(P)^{-1})
$$ with high probability, and hence, $Q_2(\{\widehat{\tau}_i\}; \{\widehat{U}_i\}) =O( \delta(P)^{-1})$ with high probability. 
%

We now show exact recovery of $\{\widehat{\tau}_i\}$. Let
 $\widehat{H}_1, \widehat{H}_2, \dots, \widehat{H}_K$ be unit norm vectors such that
$\Pi^{(\widehat{\tau}, \widehat{U})}_{k} = \widehat{H}_k \widehat{H}_k^{\top}$. Recall that $H$ is a $K \times K$ orthogonal matrix, and hence, $H_{k}^{\top} H_{\ell} = 0$ for $k \not = \ell$. 
Then for any fixed but arbitrary $\epsilon \in (0, 1/4)$, there exists with high probablity, a permutation $\sigma$ such that $H_k^{\top} \widehat{H}_{\sigma(k)} \geq 1 - \epsilon$.
Indeed, suppose that there exists a $k$ such that $H_k^{\top} \widehat{H}_{\ell} \leq 1 - \epsilon$ for all $\ell$.
Then
\begin{align*}
    Q_2(\{\widehat{\tau}_i\}; \{\widehat{U}_i\}) &= \sum_{i \colon \tau(i) = k} 
    \|(I - \Pi^{(\widehat{\tau}, \widehat{U})}_{\widehat{\tau}_i})\, \widehat{U}_i\|^2 \\ & \geq 
    \sum_{i \colon \tau(i) = k} 
    \|(I - \Pi^{(\widehat{\tau}, \widehat{U})}_{\widehat{\tau}_i})\, U_i\|^2 - 2 \sum_{i \colon \tau_i = k} \|\widehat{U} - U\|_{2 \to \infty} \|U_i\| \\
    &\geq \sum_{i \colon \tau_i = k} \|\widetilde{\theta}_i\,(I - \widehat{H}_{\widehat{\tau}_i} \widehat{H}_{\widehat{\tau}_i}^{\top})\, H_{\tau_i}\|^2 - 2\,\|\widehat{U} - U\|_{2 \to \infty} \sum_{i \colon \tau_i = k} \widetilde{\theta}_i \\
    &\geq \sum_{i \colon \tau_i = k} \widetilde{\theta_i}^2 \|H_k - \widehat{H}_{\widehat{\tau}_i} \widehat{H}_{\widehat{\tau}_i}^{\top} H_k\|^2 - 2\,\|\widehat{U} - U\|_{2 \to \infty} \sum_{i \colon \tau_i = k} \widetilde{\theta}_i 
      \\
& = \sum_{i \colon \tau_i = k} \widetilde{\theta_i}^2 (1 - (H_k^{\top} \widehat{H}_{\widehat{\tau}_i})^2) - O(\delta(P)^{-\frac{1}{2}} \log^{\frac{1}{2}}{n}) \\
& \geq  \sum_{i \colon \tau_i = k} \epsilon\,\widetilde{\theta_i}^2 - O(\delta(P)^{-\frac{1}{2}} \log^{\frac{1}{2}}{n})\gg 1/\delta(P),
\end{align*}
where the last two final inequalities follows from the fact that $\sum_{i \colon \tau_i = k} \widetilde{\theta}_i^2 = 1$ so that $$\sum_{i \colon \tau_i = k}
\widetilde{\theta}_i \leq n_k^{\frac{1}{2}} \left(\sum_{i \colon \tau_{i} = k} \widetilde{\theta}_i^2\right)^{\frac{1}{2}} = n_k^{\frac{1}{2}},$$
and hence, by Lemma~\ref{lem:estimation_U}, 
\[ \|\widehat{U} - U\|_{2 \to \infty} \sum_{i \colon \tau_i = k} \widetilde{\theta}_i \,\leq\, C\,n^{-\frac{1}{2}}\, \delta(P)^{-\frac{1}{2}} \,(\log n)^{\frac{1}{2}}\,\times\, n_k^{-\frac{1}{2}} = O(\delta(P)^{-\frac{1}{2}} \log^{\frac{1}{2}}{n}) \]
with high probability. 
The above bound for $Q_2(\{\widehat{\tau}_i\}; \{\widehat{U}_i\})$ 
contradicts the previous derivations that $Q_2(\{\widehat{\tau}_i\}; \{\widehat{U}_i\}) = O(\delta(P)^{-1})$ with high probability. 

Therefore, for any $H_k$ there must exists some $\widehat{H}_{\ell}$ such that $H_k^{\top} \widehat{H}_{\ell} \geq 1 - \epsilon$. Next, note that for any $k \not = k'$, there does not exists a $\widehat{H}_{\ell}$ such that both $H_k^{\top} \widehat{H}_{\ell} \geq 1 - \epsilon$ and $H_{k'}^{\top} \widehat{H}_{\ell} \geq 1 - \epsilon$, as otherwise
$$\|H_{k} - H_{k'}\| \,=\, \sqrt{2} \,\geq\, 2 \sqrt{2\,\epsilon} \,>\, \|H_k - \widehat{H}_{\ell}\| + \|\widehat{H}_{\ell} - H_{k'}\|,$$
which is impossible. 
Therefore, there must exist a unique bijection $\sigma$ from
$[K]$ to $[K]$, i.e., a permutation, such that $H_k^{\top} \widehat{H}_{\sigma(k)} \geq 1 - \epsilon$. 

Now for any $\widehat{U}_i$, let us do a post-processing step, if necessary, where we assign $\widehat{U}_i$
to the cluster $\ell$ for which $\|(I - \widehat{H}_{\ell} \widehat{H}_{\ell}^{\top})\, \widehat{U}_i\|$ is minimized. 
It is then easy to see that if $\tau_i = k$, then $\ell = \sigma(k)$ is the unique assignment,
provided that $\|U_i\| = \widetilde{\theta}_i = \omega(n^{-\frac{1}{2}} \delta(P)^{-\frac{1}{2}} \log^{\frac{1}{2}}{n})$, which always hold under our assumption that $\theta_{\max}/\theta_{\min} = O(1)$.
More specifically, if $\theta_{\max}/\theta_{\min} = O(1)$, then $\widetilde{\theta}_i = \Omega(n^{-\frac{1}{2}})$ for all $i$ and hence, for any $\ell$ we have
\begin{equation*}
    \begin{split}
        \|(I - \widehat{H}_{\ell} \widehat{H}_{\ell}^{\top})\, \widehat{U}_i\| 
        \, &=\, \|(I - \widehat{H}_{\ell} \widehat{H}_{\ell}^{\top})\, U_i\| \pm O(\|\widehat{U} - U\|_{2 \to \infty}) \\[.1cm]
        \, &=\, \widetilde{\theta}_i\, \|H_{\tau_i} - \widehat{H}_{\ell} \widehat{H}_{\ell}^{\top} H_{\tau_i}\| \,\pm\, O(\|\widehat{U} - U\|_{2 \to \infty}) \\[.1cm]
        \, &=\, \widetilde{\theta}_i\, \bigl(1 - (H_{\tau_i}^{\top} \widehat{H}_{\ell})^2\bigr)^{\frac{1}{2}} \,\pm\, O(\|\widehat{U} - U\|_{2 \to \infty}).
    \end{split}
\end{equation*}
The minimizer of $\|(I - \widehat{H}_{\ell} \widehat{H}_{\ell}^{\top})\, \widehat{U}_i\|$ over $\ell \in [K]$ is thus the same as the 
maximizer of $H_{k}^{\top} \widehat{H}_{\ell}$ over $\ell \in [K]$ which is given by $\sigma(k)$.
In summary, minimization of the objective function $Q_2$ yields an exact recovery of $\tau$. 
\end{proof}

\section{Proof of Theorem \ref{thm:q2_pabm_old}}

\begin{proof}
We prove that
\begin{equation}
    \sum_{i=1}^n \|(I - {\Pi}_{t_i}^{(t,\widehat{\widetilde{U}})}) \widetilde{U}_i\|^2\gg 1/\delta(P),\ \text{for any }\{t_i\}.
    \label{q2txcond}
\end{equation}
If \eqref{q2txcond} is true, we can derive
\begin{align*}
    &\,\sum_{i=1}^n \|(I - {\Pi}_{t_i}^{(t,\widehat{\widetilde{U}})}) \widetilde{U}_i\|^2
    \leq 2\sum_{i=1}^n \|(I - {\Pi}_{t_i}^{(t,\widehat{\widetilde{U}})})(\widetilde{U}_i - \widehat{\widetilde{U}}_i)\|^2 + 2\sum_{i=1}^n \|(I - {\Pi}_{t_i}^{(t,\widehat{\widetilde{U}})})\widehat{\widetilde{U}}_i\|^2\\[.1cm]
    \leq&\, 2\,\|\widehat{\widetilde{U}}-\widetilde{U}\|_{F}^2 + 2\,Q_2(\{t_i\};\{\widehat{\widetilde{U}}_i\})\leq 2\,\|\widehat{U}-U\|_{F}^2 + 2\,Q_2(\{t_i\};\{\widehat{\widetilde{U}}_i\}).
\end{align*}
Therefore, by Lemma \ref{lem:estimation_U}, we have that $Q_2(\{t_i\};\{\widehat{\widetilde{U}}_i\})\gg 1/\delta(P)$ with high probability, under Assumptions \textbf{B1} and \textbf{B2}.

What remains to show is that \eqref{q2txcond} holds. Suppose that \eqref{q2txcond} does not hold, that is, $$\sum_{i=1}^n \|(I - {\Pi}_{t_i}^{(t,\widehat{\widetilde{U}})}) \widetilde{U}_i\|^2 < C/\delta(P),$$
for some $C>0$.
We define a map $\sigma:[K]\mapsto[K]$ such that 
$$\sigma(k)=\underset{\ell\,\in\,[K]}{\arg\max}\ \#\{i:\tau_i=k,t_i = \ell\}.$$
Then,
\begin{align*}
    \sum_{i=1}^n \|(I - {\Pi}_{t_i}^{(t,\widehat{\widetilde{U}})}) \widetilde{U}_i\|^2 
    =\sum_{k=1}^K \sum_{i\colon\tau_i=k} \|(I - {\Pi}_{t_i}^{(t,\widehat{\widetilde{U}})}) \widetilde{U}_i\|^2
\geq&\,\sum_{k=1}^K \sum_{i\colon\tau_i=k,\,t_i=\sigma(k)} \|(I - {\Pi}_{\sigma(k)}^{(t,\widehat{\widetilde{U}})}) \widetilde{U}_i\|^2.
\end{align*}
Therefore,
\begin{equation}
    \sum_{k=1}^K \sum_{i\colon\tau_i=k,\,t_i=\sigma(k)} \|(I - {\Pi}_{\sigma(k)}^{(t,\widehat{\widetilde{U}})}) \widetilde{U}_i\|^2< C/\delta(P).
    \label{contradiction}
\end{equation}

The set $\{i\colon\tau_i=k,\,t_i=\sigma(k)\}$ contains at least $n_k/K$ elements by the pigeon-hole principle.
Noting that ${\Pi}_{\sigma(k)}^{(t,\widehat{\widetilde{U}})})$ is a rank-1 projection matrix, \eqref{contradiction} can not hold under the condition \eqref{pabmcond_2}.
Hence,
$$Q_2(\{t_i\};\{\widehat{\widetilde{U}}_i\})\gg1/\delta(P)\ \text{with high probability.}$$

\end{proof}

\section{Proof of Theorem \ref{thm:q3_pabm}}

\begin{proof}
We prove Theorem \ref{thm:q3_pabm} for a global minimizer $\{\widehat\tau_i\}$ of $Q_3(\{t_i\};\{\widehat{U}_i\})$.
We first show that $Q_3(\{\widehat{\tau}_i\};\{\widehat{U}_i\})=O(\delta(P)^{-1})$ with high probability, and this result, combined with the concentration of the latent positions $\{\widehat{U}_i\}$'s, ensures the exact recovery of $\{\widehat{\tau}_i\}$.
For a $(1+\epsilon)$-approximation solution $\{\widehat{\tau}_i\}$, the proof can be carried out in the same manner, since $Q_3(\{\widehat{\tau}_i\};\{\widehat{U}_i\})=O(\delta(P)^{-1})$ still holds with high probability, by definition in Eq. \eqref{main.apx.sol.3}.

The objective function $Q_3$ is defined as
$$ Q_3(\{t_i\};\{\widehat{U}_i\})=\sum_{k=1}^K \sum_{i \colon t_i={k}} \|(I - {\Pi}_{k}^{(t,\widehat{U})})\,
\widehat{U}_i\|^2.$$ 
As $\{\widehat{\tau}_i\}$ is the global minimizer of $Q_3$,
we have $Q_3(\{\widehat{\tau}_i\};\{\widehat{U}_i\})\leq Q_3(\{{\tau}_i\};\{\widehat{U}_i\}).$
We now show that $Q_3(\{\tau_i\};\{\widehat{U}_i\})$ is small. More specifically,
\begin{align*}
  \,Q_3(\{\tau_i\};\{\widehat{U}_i\})&=\sum_{k=1}^K \sum_{i\colon{\tau}_i={k}} \|(I - {\Pi}_{k}^{(\tau,\widehat{U})})\,
    \widehat{U}_i\|^2\\
    &\leq  2\sum_{k=1}^K \sum_{i\colon{\tau}_i={k}} \|(I - \Pi_{k}^{(\tau,\widehat{U})})\, (\widehat{U}_i-U_i)\|^2+2\sum_{k=1}^K \sum_{i\colon{\tau}_i=k} \|(\Gamma_{k} - \Pi_{k}^{(\tau,\widehat{U})})\, U_i\|^2\\
    &\leq  2\, \|\widehat{U}-U\|_F^2+2\sum_{{k} }\sum_{i\colon{\tau}_i={k}}\|U_i\|^2\, \|\Pi_{k}^{(\tau,\widehat{U})}-\Gamma_{k}\|^2\\
    &\leq  2\, \|\widehat{U}-U\|_F^2+2\sum_{{k} } \|\Pi_{k}^{(\tau,\widehat{U})}-\Gamma_{k}\|^2 \sum_{i\colon{\tau}_i={k}}\|U_i\|^2 \\
    &\leq  2\, \|\widehat{U}-U\|_F^2+ 2 K \sum_{k=1}^K \|\Pi_{k}^{(\tau,\widehat{U})}-\Gamma_{k}\|^2,
\end{align*}
where the final inequality follows from the fact that $\sum_{i \colon \tau_i =k} \|U_i\|^2 = K$ for all $k$ (see the discussion before Assumption~\textbf{B1}). 
Now fix a $k \in [K]$ and let $M_{k}$ be a matrix whose columns are all of the $U_i$'s
for which $\tau_i= k$. Similarly, let $\widehat{M}_{k}$ be the matrix whose columns are the $\widehat{U}_i$'s with $\tau_i={k}$. Note that $\Pi_{k}^{(\tau,\widehat{U})}$ corresponds to the
projection onto the $K$ leading left singular vectors of $\widehat{M}_{k}$. 
Once again, using the Wedin $\sin$-$\Theta$ theorem we have
$$\|\Pi_{k}^{(\tau,\widehat{U})}-\Gamma_{k} \|\leq \frac{\|\widehat{M}_{k}- M_{k}\|}{\|M_{k}\|}\leq
\|\widehat{M}_{k}-M_{k}\|$$
as $\|M_{k}\| = \|Z^{(k)}\| = 1$. Hence
$\sum_{k=1}^K \|\Pi_{k}^{(\tau,\widehat{U})}-\Gamma_{k} \|^2 \leq \sum_{k=1}^K \|\widehat{M}_{k}-M_{k}\|^2 \leq \|\widehat{U} - U\|_{F}^2$.
Combining the above bounds together with Lemma \ref{lem:estimation_U}, we obtain
\begin{equation}
  Q_3(\{\widehat\tau_i\}; \{\widehat{U}_i\}) \,\leq\, Q_3(\{\tau_i\}; \{\widehat{U}_i\}) \,\leq\, 2\,(K+1)\, \|\widehat{U} - U\|_{F}^2 \,=\, O(\delta(P)^{-1})
  \label{eq:q3_exact1}
\end{equation}
with high probability, under Assumptions \textbf{B1} and \textbf{B2}.

We now show exact recovery of $\{\widehat{\tau}_i\}$. 
Let $\Gamma_{k}$ be the $K^2 \times K^2$ orthogonal projection matrix onto the rowspace of
$Z^{(k)}$ for each $k \in [K]$, i.e., $\Gamma_{k}$ is the orthogonal projection onto the subspace spanned by the vectors $\{U_i \colon \tau_i = k\}$. We then have, from the block diagonal form for $Z$, that
$\Gamma_k$ is a $K^2 \times K^2$ diagonal matrix with diagonal entries $(\Gamma_{k})_{ss} = 1$ for $(k-1)K + 1 \leq s \leq kK$ and $(\Gamma_{k})_{ss} = 0$ otherwise. Note that $\Psi_{k}(\Gamma_{k}) = I_{K}$ for all $k \in [K]$
and $\Psi_{\ell}(\Gamma_{k}) = 0$ for all $\ell \in [K], \ell \not = k$. 
Let $\widehat{\Gamma}_1, \dots, \widehat{\Gamma}_{K}$ be the projection matrices corresponding to $\{\widehat{\tau}_i\}$, i.e., $\widehat{\Gamma}_{k} = \Pi^{(\widehat{\tau}, \widehat{U})}_k$. 

Fix $\epsilon \in (0, 1/4)$.
We show that for any $k\in[K]$, there exists $\ell\in[K]$ such that $\|I_K - \Psi_{k}(\widehat{\Gamma}_{\ell})\|_*
\leq \epsilon$, where 
$\|\cdot\|_*$ denotes the nuclear norm for matrices.  
Suppose to the contrary that this is not the case
, i.e., there exists $k\in[K]$ such that, for all $\ell \in [K]$, we have $\|I - \Psi_{k}(\widehat{\Gamma}_{\ell}) \|_* > \epsilon$. 
We then have
\begin{equation*}
    \begin{split}
        Q_3(\{\widehat{\tau}_i\}, \{U_i\}) &\geq \sum_{i \colon \tau_i = k} \|(I - \widehat{\Gamma}_{\widehat{\tau}_i})\, U_i\|^2
        \geq \max_{\ell\,\in\,[K]} \sum_{i \colon \tau_i = k, \widehat{\tau}_i = \ell} \|(I - \widehat{\Gamma}_{\ell})\, U_i\|^2\\ 
        & \geq \max_{\ell\,\in\,[K]} \sum_{i \colon \tau_i = k, \widehat{\tau}_i = \ell} U_i^{\top} (I - \widehat{\Gamma}_{\ell})\, U_i\\
        & \geq \max_{\ell\,\in\,[K]} \sum_{i \colon \tau_i = k, \widehat{\tau}_i = \ell} \lambda_i^{\top}\,\Xi_{k}^{\frac{1}{2}}\, (I_K - \Psi_{k}(\widehat{\Gamma}_{\ell}))\, \Xi_{k}^{\frac{1}{2}}\, \lambda_i \\
        &\geq \max_{\ell\,\in\,[K]} \mathrm{tr} \left[ (I_K - \Psi_{k}(\widehat{\Gamma}_{\ell}))^{\frac{1}{2}}\, \Xi_{k}^{\frac{1}{2}}\,
        \left(\sum_{i \colon \tau_i = k, \widehat{\tau}_i = \ell} \lambda_{i} \lambda_{i}^{\top}\right)
        \Xi_{k}^{\frac{1}{2}}\, (I_K - \Psi_{k}(\widehat{\Gamma}_{\ell}))^{\frac{1}{2}}\right],
  \end{split}
\end{equation*}
where $\lambda_{i} = (\lambda_{i1}, \dots, \lambda_{iK}) \in \mathbb{R}^{K}$ is the node popularity vector for vertex $i$, and $\Xi_{k} = (\Lambda^{(k)\top} \Lambda^{(k)})^{-1}$ for all $k \in [K]$. Now by the pigeonhole-principle, we have $\max\limits_{\ell\,\in\,[K]} |\{i \colon \tau_i = k, \widehat{\tau}_{\ell} = \ell\}| \geq n_k/K$ and hence, by Assumption~\textbf{B4} we have
\begin{equation*}
  \begin{split}
  Q_3(\{\widehat{\tau}_i\}, \{U_i\}) &\,\geq\,  \gamma_2\,\frac{n_k\rho_n}{K}\,\mathrm{tr} \left[ (I_K - \Psi_{k}(\widehat{\Gamma}_{\ell_*}))^{\frac{1}{2}}\, \Xi_{k}\, (I_K - \Psi_{k}(\widehat{\Gamma}_{\ell_*}))^{\frac{1}{2}}\right], 
  \end{split}
\end{equation*}
where $\ell_*$ maximizes $|\{i \colon \tau_i = k, \widehat{\tau}_{\ell} = \ell\}|$ over all $\ell \in [K]$.
Next, by the identifiability condition $\|\Lambda\|_{2 \to \infty} = \rho_n^{\frac{1}{2}}$ (Assumption \textbf{B3}), we have
$$ \|\Lambda^{(k)\top} \Lambda^{(k)}\| \,\leq\, n_k \rho_n\, \Longrightarrow\, \Lambda^{(k)\top} \Lambda^{(k)} \,\preccurlyeq\, n_{k}\rho_n\, I_K
\,\Longrightarrow\, \Xi_k \,=\, (\Lambda^{(k)\top} \Lambda^{(k)})^{-1} \,\succcurlyeq\, (n_k\rho_n)^{-1} I_K,$$
where $\preccurlyeq$ and $\succcurlyeq$ are the Lowner positive semidefinite ordering for matrices. We thus have
\begin{align*}
    Q_3(\{\widehat{\tau}_i\}, \{U_i\}) \,\geq&\,
    \gamma_2\,\frac{n_k\rho_n}{K} \mathrm{tr} \left[(I_K - \Psi_{k}(\widehat{\Gamma}_{\ell_*}))^{\frac{1}{2}}\, \Xi_{k}\, (I_K - \Psi_{k}(\widehat{\Gamma}_{\ell_*}))^{\frac{1}{2}}\right]\\[.1cm] 
    \,\geq&\, \frac{\gamma_2}{K} \, \mathrm{tr} \left[(I_K - \Psi_{k}(\widehat{\Gamma}_{\ell_*}))\right] \geq \frac{\gamma_2\, \epsilon}{K}.
    \end{align*}
A simple application of the triangle inequality 
then shows that 
$$
Q_3(\{\widehat{\tau}_i\}, \{\widehat{U}_i\}) \,\geq\, \frac{\gamma_2\, \epsilon}{2 K} 
\,\gg\, \frac{1}{\delta(P)}
$$ with high probability,
which contradicts Eq.~\eqref{eq:q3_exact1}, proved earlier.

Therefore, for any $k \in [K]$, there must exist an $\ell\in[K]$ such that $\|I_K - \Psi_{k}(\widehat{\Gamma}_{\ell})\|_*
\leq \epsilon$. Furthermore, for any $k \not = k'$, there cannot exist a common index $\ell$ such that both 
$\|I_K - \Psi_{k'}(\widehat{\Gamma}_{\ell})\|_* \leq \epsilon$ and $\|I_K - \Psi_{k}(\widehat{\Gamma}_{\ell})\|_* \leq \epsilon$,
as then
\begin{equation}
  \label{eq:norm_hat_Gamma_1}
  \begin{split}
  K &\,=\, \|\widehat{\Gamma}_{\ell}\|_* \,=\, \mathrm{tr} [\widehat{\Gamma}_{\ell}] \,\geq\, \mathrm{tr} [\Psi_{k}(\widehat{\Gamma}_{\ell})] + \mathrm{tr} [\Psi_{k'}(\widehat{\Gamma}_{\ell})] \\ 
  & \,\geq\, 2K - \|I_K - \Psi_{k}(\widehat{\Gamma}_{\ell})\|_* -
  \|I_K - \Psi_{k'}(\widehat{\Gamma}_{\ell})\|_* \,\geq\, 2K - 2\,\epsilon,
  \end{split}
\end{equation}
which is a contradiction; note that we had assumed $\epsilon < 1/4$.
Therefore, there exists a unique permutation $\sigma$ on $[K]$ such that $\|I - \Psi_{k}(\widehat{\Gamma}_{\sigma(k)})\|_* \leq \epsilon$ for all $k$.

Finally, we do a post-processing step, if necessary, wherein for every $U_i$, we assign it to the cluster $\ell$ which minimizes $\|(I - \widehat{\Gamma}_{\ell})\, \widehat{U}_i\|$. We then have
\begin{align*}
    \|(I - \widehat{\Gamma}_{\ell})\, \widehat{U}_i\| &\,=\, \|(I - \widehat{\Gamma}_{\ell})\, U_i\| \,\pm\, O(\|\widehat{U} - U\|_{2 \to \infty})\\[.1cm]
    &\,=\, \sqrt{\lambda_i^{\top}\,\Xi_{\tau_i}^{\frac{1}{2}}\, (I - \Psi_{\tau_i}(\widehat{\Gamma}_{\ell}))\, \Xi_{\tau_i}^{\frac{1}{2}}\, \lambda_i} \,\pm\, O(\|\widehat{U} - U\|_{2 \to \infty}).
\end{align*}
Suppose that $\ell = \sigma(\tau_i)$. Then
\[\lambda_i^{\top}\, \Xi_{\tau_i}^{\frac{1}{2}}\, (I - \Psi_{\tau_i}(\widehat{\Gamma}_{\ell}))\, \Xi_{\tau_i}^{\frac{1}{2}}\, \lambda_i \,\leq\, \epsilon\, \lambda_{i}^{\top}\, \Xi_{\tau_i}\, \lambda_i \,\leq\, \frac{1}{4}\, \lambda_{i}^{\top}\, \Xi_{\tau_i}\, \lambda_i.\]
Now, consider the case where $\ell \not = \sigma(\tau_i)$. 
The argument for Eq.~\eqref{eq:norm_hat_Gamma_1} shows that, for a given $\ell$,
if $\|I - \Psi_{k}(\widehat{\Gamma}_{\ell})\|_* \leq \epsilon$ for any $k$, then for all $k' \not = k$, we have
\begin{align*}
    K = \|\widehat{\Gamma}_{\ell}\|_* \,\geq\, \|\Psi_{k}(\widehat{\Gamma}_{\ell})\|_* + \|\Psi_{k'}(\widehat{\Gamma}_{\ell})\|_*
  &\,\geq\, K - \|I - \Psi_{k}(\widehat{\Gamma}_{\ell})\|_* + \|\Psi_{k'}(\widehat{\Gamma}_{\ell})\|_*\\[.1cm] 
  &\,\geq\, K - \epsilon + \|\Psi_{k'}(\widehat{\Gamma}_{\ell})\|_{*}.
\end{align*}
Hence, $\|\Psi_{k'}(\widehat{\Gamma}_{\ell})\|_* \leq \epsilon$, which also implies that $\|\Psi_{k'}(\widehat{\Gamma}_{\ell})\| \leq \epsilon$.
Then $I - \Psi_{\tau_i}(\widehat{\Gamma}_{\ell}) \succcurlyeq (1-\epsilon)\,I$ and
hence
\[
\lambda_i^{\top} \,\Xi_{\tau_i}^{\frac{1}{2}}\, (I - \Psi_{\tau_i}(\widehat{\Gamma}_{\ell}))\, \Xi_{\tau_i}^{\frac{1}{2}}\, \lambda_i \,\geq\, (1-\epsilon)\, \lambda_{i}^{\top}\,\Xi_{\tau_i}\, \lambda_i  \,\geq\, \frac{3}{4}\, \lambda_{i}^{\top}\, \Xi_{\tau_i}\, \lambda_i. 
\]
Finally, as $\lambda_i^{\top}\, \Xi_{\tau_i}\, \lambda_i\, \geq n^{-1}\, \lambda_i^{\top} \lambda_i \,\geq\, n^{-1} \gamma_1^2$ by Assumption \textbf{B3}, we have
$\sqrt{\lambda_i^{\top} \Xi_{\tau_i} \lambda_i} = \omega(\|\widehat{U} - U\|_{2 \to \infty})$ for all $i$,
which then implies
\[\argmin_{\ell\,\in\,[K]} \|(I - \widehat{\Gamma}_{\ell})\, \widehat{U}_i\| = \sigma(\tau_i)\]
for all $i$, i.e., assigning each $\widehat{U}_i$ to the cluster $\ell$ that minimizes
$\|(I - \widehat{\Gamma}_{\ell})\, \widehat{U}_i\|$ yield an exact recovery of $\tau$. 
\end{proof}

\end{document}